# Introduction to Error Correcting codes in Quantum Computers

Pedro J. Salas-Peralta
Departamento de Tecnologías Especiales Aplicadas a la Telecomunicación
Universidad Politécnica de Madrid
Ciudad Universitaria s/n, 28040 Madrid
E-mail: psalas@etsit.upm.es

**Abstract**

The goal of this paper is to review the theoretical basis for achieving a faithful quantum information transmission and processing in the presence of noise. Initially encoding and decoding, implementing gates and quantum error correction will be considered error free. Finally we will relax this non realistic assumption, introducing the quantum *fault-tolerant* concept. The existence of an error threshold permits to conclude that *there is no physical law preventing a quantum computer from being built.* An error model based on the depolarizing channel will be able to provide a simple estimation of the storage or memory computation error threshold: $\eta_{th} < 5.2 \cdot 10^{-5}$. The encoding is made by means of the [[7,1,3]] Calderbank-Shor-Steane quantum code and the Shor's fault-tolerant method to measure the stabilizer's generators is used.

**Resumen**

El objetivo de este artículo es la revisión de los fundamentos teóricos que permiten una correcta transmisión y procesado de la información cuántica en presencia de ruido. Inicialmente, los procesos de codificación, decodificación, aplicación de puertas y corrección de errores se considerarán sin error. Finalmente relajaremos esta consideración no realista, lo que conducirá al concepto de *tolerancia a fallos*. La existencia de un umbral de error permite concluir que *no hay ninguna ley física que impida construir un ordenador cuántico*. Mediante un modelo de error basado en un canal despolarizante, se hará una estimación simple para el umbral de los errores de memoria: $\eta_{th} < 5.2 \cdot 10^{-5}$. La codificación se realiza mediante un código cuántico [[7,1,3]] de Calderbank-Shor-Steane y el método de Shor tolerante a fallos para medir los generadores del estabilizador.

## 1. Introduction

Quantum Mechanics (QM) has traditionally been used to study microscopic systems, achieving unquestionable successes in such diverse fields as atom structure, elementary particles, solids, liquids, molecules, nuclei, radiation etc. It is currently extending into a field traditionally dominated by a classic description: the Computation and Information Theory. Although the devices making up a classic computer work according quantum laws, they do not make use of the quantum representation of the information, but they continue to use the classic version: bits. The recognition that the information is closely related to its physical representation and the non local character of the QM, is opening up an unsuspected perspective from a classic point of view for data processing [1]. In this context, the concept of the quantum computer appears as a device that takes advantage of the quantum evolution to obtain new forms of information processing. Its minimum unit of information is the *quantum bit or qubit*, that consists of a state (coherent superposition of another two representing the classic possibilities |0> and |1>) of the type |q> = a|0> + b|1>, where a and b are complex numbers.

Similarly to classic computers, quantum computers experience the presence of noise that induces errors into them. Unlike classic computers, the quantum ones must handle coherent superposition and entangled states, allowing interference phenomena analogous of those produced when the light crosses a system of two slits of a similar size to its wavelength. Unfortunately the superposition of states is extremely sensitive to the noise and they are easily destroyed, due to an uncontrollable interaction with the environment. This process is known as *decoherence* [2]. It would be possible to think about eliminating it by improving the isolation of the device. Nevertheless the extraction of the information at the end of any computation process, always implies some type of measurement, this is why simple isolation is not a solution. In addition, it is impossible to completely eliminate all the interactions that come from the environment. Until 1995 it was believed that the unavoidable decoherence would prevent the quantum information processing from showing its advantages with respect to the classic case. Luckily the things were going to change [3].

The objective of the present paper will be to show how the noise is not an unsolvable problem to build a quantum computer. After a brief introduction to classic error correction, the characteristics of quantum errors are introduced, and the noise effect will be exemplified by means of the Grover algorithm including five qubits. Several strategies introduced to control the decoherence will be reviewed, focusing the explanation on the quantum error correcting codes. A simple numerical method, encoding a qubit by means of the [[7,1,3]] fault tolerant quantum code, will permit us to infer the existence of an error threshold below which a sufficiently long quantum computation would be possible. Finally, concatenated codes will promise us to improve the error correction capabilities.

## 2. Classic errors and their correction

In order to understand the main ideas in quantum error correction, we start with some classic background.

Classic information is represented by means of an alphabet of p symbols. The binary alphabet (p=2) is made up of two symbols {0, 1} and the information contained in each symbol is called a binary digit or a bit. The information processing involves representing it as bit strings, sending them through a channel or carrying out a computation and, finally, arriving at a result. Unfortunately, noise can always corrupt the information. A possible strategy to preserve the classic information against the noise effect is by means of an encoding method. The information contained in a single bit is spread out along a bit string of length n, called classic register or *codeword*. From a mathematical point of view, the set of all words of length n ($V_2^n$) with the modulo 2 arithmetic, could have a structure. Of particular importance are the sets of codewords C $\subseteq$ $V_2^n$, which have a vector space structure, called *linear codes*. This structure makes the correction process easier. It is also possible to define a product operation, which together with the addition defines a finite *field* also called a *Galois field*. The binary alphabet {0, 1} is an example and will be referred as the GF(2) field or as a vector space $V_2$.

The *Hamming distance* d(u,v) between two codewords u, v $\in$ C $\subseteq$ $V_2^n$ is the number of coordinates where the vectors u and v differ:

$$d(u,v) = |\{i : 1 \leq i \leq n, \ u_i \neq v_i\}| \qquad (1)$$

The bars signify the number of elements of this set. The distance d satisfies the axioms for a metric on $V_2^n$. The *minimum distance* of a code is the smallest distance between two different codewords. The number of non-zero components of a binary string of $V_2^n$ is called *weight* (or Hamming weight, $W_H$), and the distance between u and v is $d(u,v) = W_H(u-v)$.

The code capability to correct the errors is represented by the code distance. Suppose the emitter sends the codeword $u \in C$ through a classic channel affected by some error probability, and the receiver detects a slightly different codeword u' = u+e ≠ u, affected by the error $e \in V_2^n$. By means of the minimum distance decoder, the word u' = u+e will be decoded as the closest codeword, according the Hamming distance. Having a code C with distance d ≥ 2t + 1 (or d > 2t), the receiver will recognize the correct codeword u from u' if and only if it fulfils $d(u,u') = W_H(e) \leq t$, because in this case d(u,u') < d(v,u'), ∀v ∈ C. As a consequence, the code C with distance d will correct any word u' = u+e, satisfying $W_H(e) \leq t$, and it will be a *t-error-correcting code*. Thus, good error correction means large minimum distance. On the other hand, fast transmission rate means many codewords, with small distance between them. This tension is the basis of coding theory.

To visualize the code distance and correcting capabilities, each codeword $u_j \in C$ is represented as the "centre" of a "sphere" with radius $t = \lfloor (d-1)/2 \rfloor$. The sphere contains all binary sequences $v = u_j + e \in V_2^n$ such as $d(u_j,v) \leq t$. Since the code C is t-error-correcting, the spheres are disjoint. The vectors inside the t-sphere come from $u_j$ affected by an error e of weight $W_H(e) \leq t$. Figure 1 shows the case d = 5 (t=2). Any erroneous codeword u' = $u_1 + e_1$ with $W_H(e_1) = 2$ is successfully corrected with a d = 5 code, but not if u' = $u_1 + e_2$ with $W_H(e_2) = 3$. In this case u' would be wrongly corrected as $u_2$.

If C is a vector subspace of $V_2^n$, d is the smallest weight of a non-zero codeword. As a consequence, a binary classic code of dimension k (including $2^k$ codewords) of length n and minimum distance d is noted as $C = [n,k,d] \subseteq V_2^n$. A linear code [n,k,d] (i.e. a linear subspace) can be specified in either of two ways:

1) The k basis vectors of C are arranged in the k×n *generator matrix G*. Thus

$$C = \{xG, \ x \in V_2^k\} \quad (2)$$

this is useful for encoding. If the messages to be transmitted are all k-tuples x over $V_2$, then we can encode them as the codewords xG.

2) It is possible to define a scalar (or inner) product in $V_2^n$ as the standard rule of multiplying the components and making the addition modulo 2. Two vectors are orthogonal if their scalar product is zero. The code can also be determined as the subspace orthogonal to some predetermined set of vectors. Each orthogonality condition divides the space by two, and then we can specify a code having a $2^k$ vectors (and dimension k) through its orthogonality to (n-k) vectors. These vectors can be arranged as a (n-k)×n matrix called *parity-check matrix* $H_C$, and the code can be specified as

$$C = \{v \in V_2^n, \quad H_C v^T = 0\} \quad (3)$$

This is useful for error correction. The set of correctable errors S must satisfy $\forall e_i, e_k \in S \subseteq V_2^n$, ∀ u, v ∈ C, if u ≠ v then $u + e_i \neq v + e_k$. If the vector $u + e_i$ is detected, the receiver can correctly infer the codeword u. This process is very easy for linear codes using the parity check matrix. Suppose the receiver detects the vector u + e with u ∈ C and e ∈ S. Applying the parity-check matrix $H_C (u + e)^T = H_C u^T + H_C e^T = H_C e^T$. The vector $H_C\, e^T = s \neq 0$, having (n-k) components, characterizes the error, it is called the *error syndrome* and does not depend on u. Because the total number of syndromes is $2^{n-k}$, the code can correct the same number of different errors. If we can deduce the error e from its syndrome, the correction is immediate.

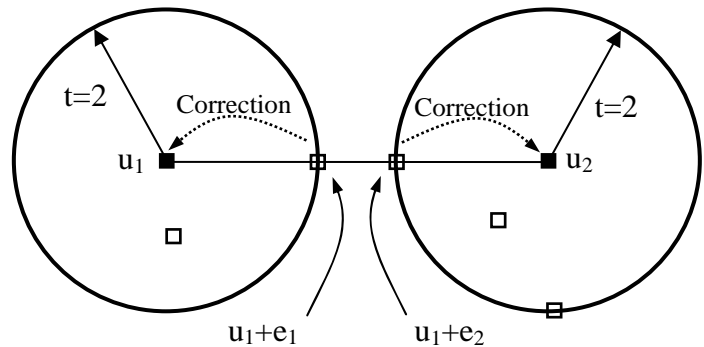

Fig 1. Geometrical representation of a classic code with distance 5. Each codeword $u_j$ (black squares) is at the "centre" of a "sphere" with radius t = 2. An erroneous codeword (unfilled square) u' = $u_1$ + $e_1$ with $W_H(e_1)$ = 2 is successfully corrected as $u_1$. If $W_H(e_2)$ = 3 the codeword u' = $u_1$ + $e_2$ is wrongly corrected as $u_2$.

Even though a classic code is not necessarily a vector space, in this paper we will be concerned only with linear codes. A simple classic code is the repetition code in which the 0 bit is encoded copying the bit three times as a codeword (000) and the bit 1 is encoded as the string (111). The set of all codewords of length three span a vector space and the set {(000), (111)} is a basis of a dimension two subspace $C \subseteq V_2^3$. This subspace C is our classic repetition code of length three. The code {(000), (111)} can be specified as the subspace orthogonal to (110) and to (101), and both vectors written as a 2×3 matrix is the *parity-check matrix* $H_C$, and the code satisfies the condition $\forall u \in C$, $H_C u^T = 0$. The 1×3 *generation matrix* G is (111). Clearly the code has distance three, so is written as [3,1,3].

If we want to send a bit 0 through a noisy channel, using the repetition code, we send (000). Classic noise appears as bit-flip errors and can be represented as error codewords of $V_2^3$. If the channel introduces a bit-flip error (with a probability ε) into the third bit, e = (001), it will be enough for the receiver to watch the three bits, and finding the syndrome (01), it will suppose that an error in the third copy has occurred, recovering the bit to replace the (000) (*majority voting decoding*). For this method to be advantageous, it is necessary for the probability of correct transmission (1-ε) of each bit to be higher than 50%, otherwise the majority voting method would provide an erroneous answer. A wrong decoding will occur if the received word has two 1's.

Given a parity-check matrix, each of its columns represents the syndrome for an error. If all the columns are different, the code can correct one bit-flip and is called *Hamming code* whose general parameters are [$2^r$-1, $2^r$-1-r, d] with r ≥ 2. An example that will be used in the quantum construction is the [7,4,3]. This code has a subcode $C^\perp \subset C$, whose codewords of even weight are orthogonal (with respect to the scalar product) to those of C. In general, given a code C = [n,k,d] its orthogonal or *dual* code is $C^\perp$ = [n, n-k, $d^\perp$] and if $C \subset C^\perp$ it is said that C is *weakly self-dual*, and if C = $C^\perp$, C is *self-dual*. The property of weakly self-duality will be used in the quantum error correcting code construction. Besides the Hamming codes,

Reed-Muller codes are an interesting family of *weakly self-dual* and *self-dual* codes. Its parameters are:

$$RM(r,m) = \left[ n = 2^m, \; k = \binom{m}{0} + \binom{m}{1} + \cdots + \binom{m}{r}, \; 2^{m-r} \right] \quad (4)$$

with $0 \leq r \leq m$.

Other classic codes can be created be means of different scalar products and higher alphabet dimensions.

There are several bounds concerning classic codes. One of them is the Hamming bound reflecting that a code C = [n,k,d] with block length n, can correct errors of weight t if there is enough room in the total vector space (of dimension n) to accommodate the errors:

$$\text{Number of different errors} = \sum_{i=1}^{t} \binom{n}{i} \leq 2^{n-k} = \text{total number of different syndromes} \quad (5)$$

Let the codewords be $\{u_i, i=1,\ldots,2^k\}$. For each codeword we can draw a "sphere" with "centre" at $u_j$ and "radius" t. The sphere contains all binary sequences v such as $d(u_j,v) \leq t$. Since the code C is t-error-correcting, the spheres are disjoint. The summation in equation (5) is the number of $v = u_j + e$ vectors inside the t-sphere coming from $u_j$ affected by an error e of weight $W_H(e) \leq t$. In order to differentiate errors, this value must be smaller than the number of different syndromes. A code is *perfect* if it attains the equality in (5), and the union of all the spheres is $V_2^n$.

**3. Origin of the quantum errors**

All of the systems are subject to noise of diverse origin (interaction with the environment, incorrect application of gates, etc), ending up with the appearance of errors. In order to carry out a quantum computation, it is necessary to eliminate or control these errors.

Focusing on the quantum computation and from the point of view of their origin, these errors can be *internal* and *external* (figure 2). The internal ones appear even if there is no interaction with the environment and originate in the wrong operation of some parts of the hardware. Several types of them include:

1. *Errors in the preparation of the initial states.*

   Classically the errors appearing in the preparation of the initial state, propagate exponentially with respect to the number of steps, nevertheless, from a quantum point of view, they are constant. Let us suppose that we prepare an initial state $|\psi_i\rangle$ evolving by means of a process characterised by a Hamiltonian $\hat{H}$ (or an evolution operator $\hat{U} = e^{-i\hat{H}t}$, $h/2\pi=1$) until the final state $|\psi_f\rangle$. In the case of a perfect preparation:

$$|\psi_i\rangle \rightarrow |\psi_f\rangle = \hat{U}(t)|\psi_i\rangle = e^{-i\hat{H}t}|\psi_i\rangle \quad (6)$$

   If the initial state corresponds to a set of single qubits all of them in the state $|0\rangle$ except the k qubit having an error $\varepsilon$:

$$|\psi_i\rangle = |0\rangle \otimes |0\rangle \otimes \ldots \otimes \left(\sqrt{1-\varepsilon^2}|0_k\rangle + \varepsilon|1_k\rangle\right) \otimes \ldots \otimes |0\rangle = \sqrt{1-\varepsilon^2}|\psi_i\rangle + \varepsilon|\text{waste}\rangle \quad (7a)$$

and its time evolution will be:

$$|\psi_f\rangle = \sqrt{1-\varepsilon^2}\,\hat{U}|\psi_i\rangle + \varepsilon\hat{U}|\text{waste}\rangle = \sqrt{1-\varepsilon^2}\,|\psi_f\rangle + \varepsilon|\text{dirty waste}\rangle \qquad (7b)$$

which implies that the initial error (given by $|\varepsilon|^2$ in $|\psi_i\rangle$) does not increase in the evolution. This behaviour arises from the linearity of the QM. In some cases, the quantum algorithms are even sensitive to these errors in the amplitude, and its accumulation become dangerous. It is necessary to pay special attention when the initial errors affect, not the amplitude but the relative phases [4], whose effect depends on the considered quantum algorithm.

2. *Hardware errors.*
   Their origin is in the noisy gate application, especially when they are analogical (working with continuous parameters) and can be described as *unitary errors* due to an error term $\hat{\eta}$ in the noiseless Hamiltonian $\hat{H}_0$: $\hat{H}_\eta = \hat{\eta} + \hat{H}_0$. The noiseless evolution is $e^{-i\hat{H}_0 t}|\psi_i\rangle = |\psi_f\rangle$. If the error operator $\hat{\eta}$ is small enough, $[\hat{H}_0, \hat{\eta}] \cong 0$ and the $\hat{\eta}$ effect on $|\psi_i\rangle$ is $e^{-i(\hat{\eta}+\hat{H}_0)t}|\psi_i\rangle = e^{-i\hat{\eta}t}|\psi_f\rangle$. The exponential can be expanded and only retain the linear term, and $|\psi_i\rangle$ evolves to $(1 - i\hat{\eta}t)|\psi_f\rangle$. So the error probability grows quadratic in time.

3. *Read-out errors of the results at the end of the process.*
   Concerning the amplification of the results from the quantum domain until the classic macroworld.

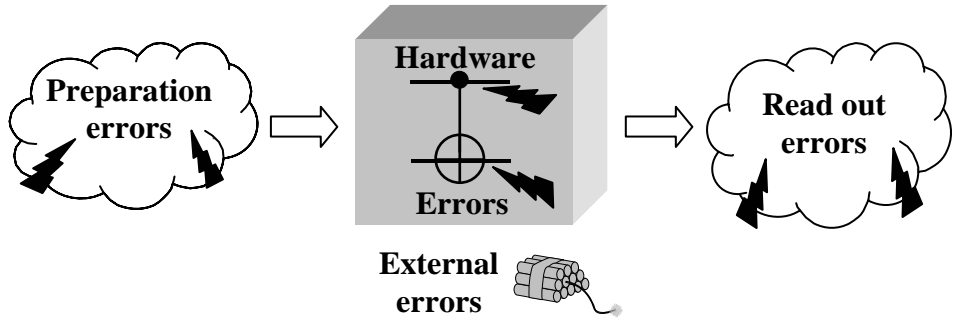

Fig. 2. Framework of the different error sources in a quantum computer.

In addition to the internal errors, external ones may appear because the system is not completely isolated from its environment, leading to a decoherence, and giving rise to a *non-unitary* evolution of the states in the quantum computer. This loss of coherence is the most serious problem which the future quantum computers face.

## 4. Problems in the correction of quantum errors

At the time of designing methods to control quantum errors, the following question arises; can we apply classic strategies to the quantum systems? For example, could classic error correcting codes be used? The answer to this question has been negative by the following problems:

*1. Continuous errors.*

Classically, the only permissible errors are those of bit-flip (transformation of a bit 0 to 1 or the reverse) and are *discrete*, but for the quantum the situation is more complicated. The errors can affect the modules of the coefficients a and b in the qubit superposition (amplitude decoherence), as well as its relative phases (phase decoherence), both being *continuous* ones. For instance, if the physical representation of qubits implies that |0> is the fundamental state of an atom, whereas |1> corresponds to an excited state, a spontaneous decaying process produces an amplitude decoherence. Its time evolution will be:

$$|q(t)\rangle = \frac{1}{\sqrt{|a|^2 + |b|^2 e^{-2\gamma t}}} \left\{ a|0\rangle + be^{-\gamma t}|1\rangle \right\} \tag{8}$$

In the case where it only affects its relative phase, the qubit is transformed into (a|0> + $be^{i\phi}$|1>). If $\phi = \pi$ we have a discrete phase-flip error, analogous to the classic bit-flip. The phase-flip is only a quantum error.

*2. Impossibility of introducing redundant information copying it.*

One of the ideas on which the correct transmission of classic information is based, is the possibility of copying it (introducing redundancy), which allows information recovery in the presence of noise as indicated in section 2.

Unfortunately, quantum mechanically it is not possible to copy unknown qubits perfectly due to the *impossibility of cloning unknown qubits* [5]. In order to copy a qubit we need to know about it. Given a qubit |q> = a|0> + b|1> (with unknown coefficients a and b), we would have to measure it to acquire the a and b values, but in doing so we would produce its collapse, destroying it irreversibly.

*3. Measurement problem.*

In order to correct the errors, we must measure the state of the system (for example some qubits) to find out what type of error has occurred. When doing it the state collapses with the consequent irreversible loss of information.

In the following we will review the way in which all these problems were solved.

**5. Discretization of quantum errors**

In 1995 the way was discovered to transform typically continuous quantum errors, in discrete solving the first aforementioned problem. The strategy consists of embedding the {environment + qubit} continuous evolution only in the first, making a discrete description of the qubit state evolution. Formally the interaction process of a qubit with its environment can be described by means of the following evolution [6]:

$$|0\rangle|e\rangle \xrightarrow{\hat{U}(t)} c_{00}|e_0\rangle|0\rangle + c_{01}|e_1\rangle|1\rangle$$
$$|1\rangle|e\rangle \xrightarrow{\hat{U}(t)} c_{10}|e_0\rangle|0\rangle + c_{11}|e_1\rangle|1\rangle \tag{9}$$

{|0>, |1>} being the qubit states and |e> the initial state of the environment. The total initial state is the tensor product of the qubit and the environment states, and evolve (unitarily) by

means of the coefficients $c_{ij}$ that depend on the noise. This is the most general form of the noise effect, assuming that qubits do not leave the dimension two {|0>, |1>} subspace of the total Hilbert space $\mathcal{H}_2$.

The qubit evolution whose initial (t=0) state is $|q(0)> = a|0> + b|1>$ can be expressed as:

$$|q(0)\rangle|e\rangle = (a|0\rangle + b|1\rangle)|e\rangle \xrightarrow{\hat{U}(t)} |\psi(t)\rangle = \{|e_I\rangle\hat{I} + |e_X\rangle\hat{X} + |e_Y\rangle\hat{Y} + |e_Z\rangle\hat{Z}\}|q(0)\rangle \quad (10)$$

The states $|e_i\rangle$ describe the environment, and $\hat{I}, \hat{X}, \hat{Y}, \hat{Z}$ are the operators whose representation in terms of the Pauli matrices $\{I, \sigma_X, \sigma_Y, \sigma_Z\}$ is:

$$\hat{I} \equiv \begin{pmatrix} 1 & 0 \\ 0 & 1 \end{pmatrix} \quad \hat{X} \equiv \begin{pmatrix} 0 & 1 \\ 1 & 0 \end{pmatrix} = \sigma_X \quad \hat{Y} \equiv \begin{pmatrix} 0 & -1 \\ 1 & 0 \end{pmatrix} = -i\sigma_Y \quad \hat{Z} \equiv \begin{pmatrix} 1 & 0 \\ 0 & -1 \end{pmatrix} = \sigma_Z \quad (11)$$

sometimes called *canonical set of errors*, whereas the states of the environment are:

$$|e_I\rangle = \frac{1}{2}(c_{00}|e_0\rangle + c_{11}|e_1\rangle) \qquad |e_X\rangle = \frac{1}{2}(c_{10}|e_0\rangle + c_{01}|e_1\rangle)$$
$$|e_Y\rangle = \frac{1}{2}(c_{01}|e_1\rangle - c_{10}|e_0\rangle) \qquad |e_Z\rangle = \frac{1}{2}(c_{00}|e_0\rangle - c_{11}|e_1\rangle) \quad (12)$$

The state $|\psi(t)\rangle$ reflects a correlation between the states of the environment and those of the qubit, describing a mixed state that has lost some coherence. If we could make a measurement on the joint state vector $|\psi(t)\rangle$ of the {environment + qubit} *conserving the qubit coherence*, we would collapse the state in one of the following terms:

$$|\psi(t)\rangle \xrightarrow{\text{Measure}} \begin{cases} |e_I\rangle\hat{I}|q(0)\rangle = |e_I\rangle\{a|0\rangle + b|1\rangle\} & \rightarrow \text{State without error} \\ |e_X\rangle\hat{X}|q(0)\rangle = |e_X\rangle\{a|1\rangle + b|0\rangle\} & \rightarrow \text{Bit-flip error} \\ |e_Z\rangle\hat{Z}|q(0)\rangle = |e_Z\rangle\{a|0\rangle - b|1\rangle\} & \rightarrow \text{Phase-flip error} \\ |e_Y\rangle\hat{Y}|q(0)\rangle = |e_Y\rangle\{a|1\rangle - b|0\rangle\} & \rightarrow \text{Phase and bit-flip error} \end{cases} \quad (13)$$

with a collapse probability given by $|\varepsilon_i|^2 = |\langle q(0), e_i|(\hat{A}_i^+ \otimes \hat{I})\hat{U}(t)|q(0), e\rangle|^2$ and $\hat{A}_i \in \{\hat{I}, \hat{X}, \hat{Y}, \hat{Z}\}$. Note that $|\varepsilon_i|^2$ imply the overlap between the environment states (generally neither orthogonal nor normalised), and their value can depend on time by means of $\hat{U}(t)$. The process (13) has a fundamental importance for several reasons:

a) The complete qubit evolution can be expressed by means of four basic operators, providing a *discrete* translation of the noise effect. It could be said that the qubit evolution is represented via three errors: bit-flip ($\hat{X}$), phase-flip ($\hat{Z}$) and both jointly ($\hat{Y}$). This fact shows that the matrices are a base for the 2×2 matrices. For the same reason the errors coming from unitary evolutions, can be interpreted in this form, being able to work without the environment states explicitly. In fact, for that the error identification to be complete, the environment states must be orthogonal.

b) The noise is *independent* of the qubit state considered, which allows its initial coherence to be conserved after the measurement step.

c) This state is the front door to the error correction process. If we have some way of recognising which state we have obtained by measuring |ψ(t)>, the error correction is immediate, simply applying the inverse transformation of the detected error, since they are unitary.

## 6. Independent Error model.

The classic error model (or channel) par excellence, considers the errors in different bits as independent. Even if this model does not adjust absolutely to the reality, it can provide some valuable consequences.

In QM is possible to introduce an analogous noisy channel called a *depolarising error model*, in which the environment states $\{|e_i\rangle, i=I,X,Y,Z\}$ are orthogonal and its scalar product is $|\langle e_i|e_j\rangle|^2 = \delta_{ij}\varepsilon/3$ $(i,j\neq I)$, where $\varepsilon/3$ is the probability (constant) of one of the three possible errors taking place, whereas the probability of no error is $|\langle e_I|e_I\rangle|^2 = (1-\varepsilon)$. The qubit evolution can be represented by means of the operator $\hat{U}_D$:

$$\hat{U}_D\left(|q(0)\rangle \otimes |e\rangle\right) = \left\{\sqrt{(1-\varepsilon)}|e_I\rangle\hat{I} + \sqrt{\frac{\varepsilon}{3}}\left[|e_X\rangle\hat{X} + |e_Y\rangle\hat{Y} + |e_Z\rangle\hat{Z}\right]\right\}|q(0)\rangle \qquad (14)$$

The error model is not completely unrealistic if one assumes that single qubits are located at well separated spatial positions, as in an ion-trap realization of a quantum computer.

As much as we are interested in handling and transmitting quantum information just as if we consider the possibility of some type of encoding, we will handle sets of n qubits called *quantum registers* $|q_1 q_2 \ldots q_n\rangle$. To see how the decoherence affects the registers, we can make some hypotheses about the error model to simplify the problem and constitute an approach to the reality [7]:

1. Errors *locally independent*.
   If the environments to which the qubits are connected (at the same time step) are different and not correlated, the errors in different qubits will be independent.

2. Errors *sequentially independent*.
   The errors in same qubit during different time steps are not correlated.

3. We assume a *small interaction qubit-environment*.

4. *Error-scalability independence*.
   The qubit error probability is independent of the number of them which are used in the computation.

Under these hypotheses, errors that affect an increasing number of qubits are less probable, and the error operators for an n-qubit register are the tensor product of those one qubit operators:

$$\hat{A}_{\{i_1,i_2,\ldots,i_n\}} = \hat{A}^1_{i_1} \otimes \hat{A}^2_{i_2} \otimes \ldots \otimes \hat{A}^n_{i_n} \qquad (15)$$

where the superscript refers the qubit, and the subscript varies from 1 to 4: $\hat{A}^m_{i_m}$ (for the m qubit) $\in$ {I($i_m$=1), $\sigma_X$($i_m$=2), -i$\sigma_Y$($i_m$=3), $\sigma_Z$($i_m$=4)}. In the depolarising error model, the evolution of an n-qubit quantum register is:

$$\hat{U}_D(|q_1q_2...q_n\rangle|e\rangle) = |\Psi(t)\rangle =$$

$$= \left\{(1-\varepsilon)^{n/2}(\hat{I}_1 \otimes ... \otimes \hat{I}_n)|e_0\rangle + (1-\varepsilon)^{(n-1)/2}\sqrt{\frac{\varepsilon}{3}}\sum_{i=2,3,4}\left\{\hat{A}_i \otimes \hat{I}_2 \otimes ... \otimes \hat{I}_n|e_i^1\rangle + ... + \right.\right.$$

$$\left.\left.+ \hat{I}_1 \otimes ... \otimes \hat{I}_{n-1} \otimes \hat{A}_i|e_i^n\rangle\right\} + ... + \left(\frac{\varepsilon}{3}\right)^{n/2}\sum_{i_1,i_2,...,i_n=2,3,4}(\hat{A}^1_{i_1} \otimes ... \otimes \hat{A}^n_{i_n})|e^{1,...,n}_{i_1...i_n}\rangle\right\}|q_1q_2...q_n\rangle \quad (16)$$

As the interaction with the environment is small (hypothesis 3), the successive terms decrease quickly. A measurement of the register $|\Psi(t)\rangle$ will produce a collapse in one of its terms according to its probability. In the equation (16) each error $\hat{A}_i$ corresponds to three terms $\{\hat{X}, \hat{Y}, \hat{Z}\}$ (the $\hat{I}$ term is explicitly shown) and the probability of an error appearing in a given qubit is $\varepsilon$, the one in which m errors appear in the register is P(n,m) = $\binom{n}{m}$ $(1-\varepsilon)^{n-m}\varepsilon^m$, describing a Bernouilli distribution of (1-$\varepsilon$) probability. If $\varepsilon$ is small enough, the term with greater collapse probability is a register without error.

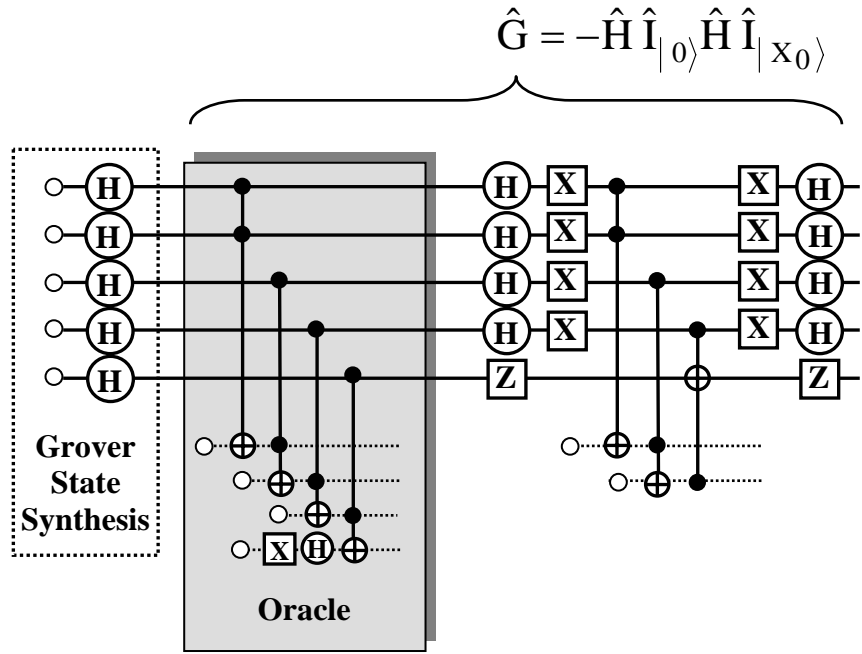

Fig. 3. Quantum Circuit implementing the Grover search algorithm for a data base with $2^5$ terms. The oracle detecting the searched state ($|X_0\rangle$ =$|11111\rangle$) is simulated by means of four Toffoli gates. Open circles represent $|0\rangle$ states.

In order to observe the destructive effect that the errors cause in the quantum algorithms (decoherence), a numerical simulation of the Grover algorithm is made. The errors are introduced by means of the depolarising error model. The free evolution (or memory) errors have a $\varepsilon/3$ probability per single qubit and time step, whereas the gates affecting single qubits have a $\gamma$ error. The CNOT gates have a $\gamma/15$ error, describing an isotropic probability for the 15 errors in the set $\{\hat{I}, \hat{X}, \hat{Y}, \hat{Z}\} \otimes \{\hat{I}, \hat{X}, \hat{Y}, \hat{Z}\}$. Toffoli gates are affected by an error probability of $\gamma/N$, where $N = 63$ are the total number of error possibilities (except one) of the set $\{\hat{I}, \hat{X}, \hat{Y}, \hat{Z}\}^{\otimes 3}$. The simulation is made by means of a Montecarlo method with a statistic greater than or equal to $100 \times \max\{1/\varepsilon, 1/\gamma\}$.

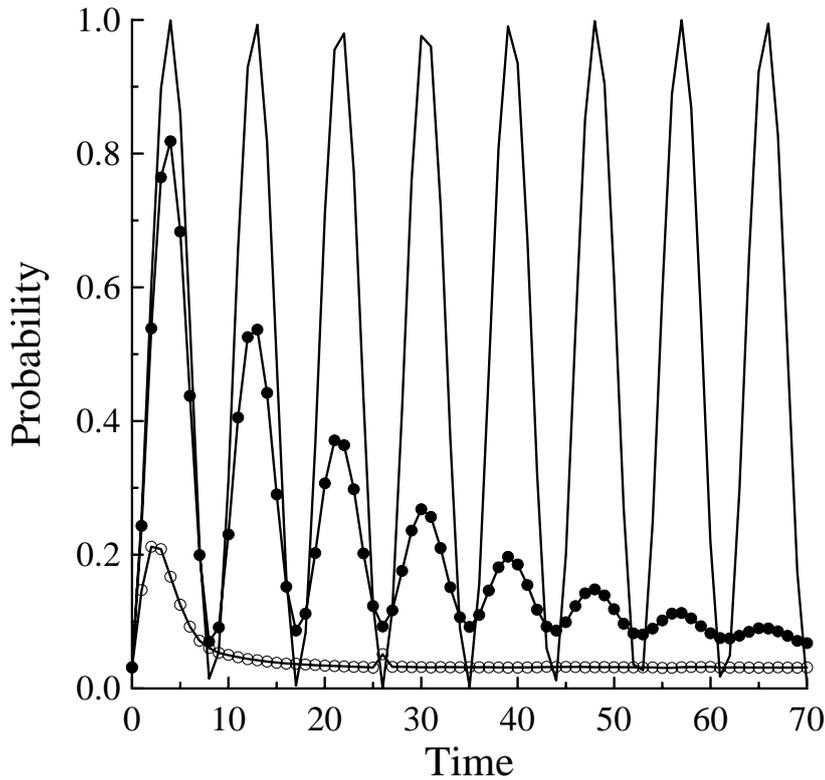

Fig. 4. Evolution of the coefficient squared (probability) for the searched state ($|11111\rangle$) versus time. Time means the number of Grover gates ($\hat{G}$) applied. Solid line represents the evolution without error; dashed lines include error: • $\varepsilon = \gamma = 0.001$ and o $\varepsilon = \gamma = 0.01$.

The Grover algorithm [8] implements the operator $\hat{G} = -\hat{H}^{\otimes n} \hat{I}_{|0\rangle} \hat{H}^{\otimes n} \hat{I}_{|X_0\rangle}$, where $\hat{H}^{\otimes n}$ is a Hadamard rotation of all the n qubits and the operators $\hat{I}_{|\phi\rangle} = \hat{I} - 2|\phi\rangle\langle\phi|$ represents inversions with respect to the state $|\phi\rangle$. The searched state is symbolized by $|X_0\rangle$, whereas $\hat{I}_{|X_0\rangle}$ represents an oracle making an inversion with respect to the searched state, acting as a black box. The simulation is made within a modest data base with $2^5 = 32$ elements. Its implementation requires at least five qubits. In the simulation the looked for element is $|X_0\rangle = |11111\rangle$ and the oracle is implemented by the quantum gate CNOT(1,...,5;6), whose control qubits are the first five qubits of Grover state and whose target is the sixth qubit in the state ($|0\rangle - |1\rangle$). The gate CNOT(1,...,5;6) is carried out [9] by means of four Toffoli gates with four

additional qubits. The operator $\hat{I}_{|0\rangle}$ is applied by means of a gate $(\hat{X}_{11111} \text{ CZ}(1,..,4;5) \hat{X}_{11111})$, between qubits of the Grover state, for which two additional qubits are needed. The simplectic notation ($\hat{X}_{11111} = \hat{X} \otimes \hat{X} \otimes \hat{X} \otimes \hat{X} \otimes \hat{X}$) will be used to express the tensor product of Pauli operators (see [3] Preskill) and CU(i;k) means a control-U gate acting on the k qubit depending on the i qubit value. The total circuit [10] for the Grover algorithm appears in figure 3.

Two calculations have been made with $\varepsilon = \gamma = 0.001, 0.01$ whose results are compared with the case in which there is no decoherence ($\varepsilon = \gamma = 0$). As it is appraised in figure 4, even for a small search such as the present one (32 elements), the error effect quickly destroys the advantages of the algorithm. Whereas for $\varepsilon = \gamma = 0.001$ the first maximum of the probability for the searched state reaches a value of 0.8, for $\varepsilon = \gamma = 0.01$ its value is only 0.2. Decoherence causes an attenuation of the Grover oscillations until the limit value of 1/32 is reached, in the long time region.

**7. Quantum strategies for error control**

Two great strategies for the error control can be implemented: *passive methods*, useful when we need a transmission of information over short distances. The most elementary are based on a complete isolation between the computer and its environment to minimise the noise. A second general method implies an *active* stabilisation (necessary in more complex processes) by means of some type of error detection and correction.

Classic deteriorated information is still recoverable if some *redundancy* has been introduced. Unfortunately, it is not possible to use this redundancy in the quantum case, due to the impossibility of cloning unknown qubits. However methods have been developed that allow us to control the qubit decoherence, thus solving the second problem settled in section 4. Next we review some of the main strategies (figure 5).

1. *Quantum error preventing codes (QEPC).*

   These codes could be described as *active* methods in the sense that they prevent the error appearance, although if these do take place they are incapable of correcting them. They are based on the quantum Zeno effect.

2. *Quantum error avoiding codes (QEAC).*

   Encode the information in states of certain subspaces that do not undergo decoherence, and they are called *decoherence free subspaces (DFS)*. Error detection is not needed and they are useful with specific types of noise.

3. *Quantum error correcting codes (QECC)*

   This is an *active* strategy defined as the pair $Q(\hat{E}, \hat{R})$, made up of an encoding operation $\hat{E}$ and a recovery method $\hat{R}$. They are methods capable of detecting and correcting quantum errors.

Notice that the corrected final system could still contain some errors, shown in figure 5 as a heavy line around the system (and a ψ somewhat deformed) that differentiates it from the initial state. The QEPC are applied *before* the errors are dangerously accumulated. On the other hand, the QEAC circumvent the problem of error appearing. Even in this case, the final state can contain errors since the symmetries upon which these methods are based can only be approximate. Finally, the QECC are even applied after the appearance of errors.

Actually, the above distinction among the different quantum codes or strategies is not as radical as it could seem. For instance a QECC applied very quickly could make the effect of a QEPC. Otherwise, some errors could not affect the encoded states of a QECC, so for these errors the code is functioning as a QEAC. In spite of that, the previous classification help us to arrange the methods used to control the decoherence.

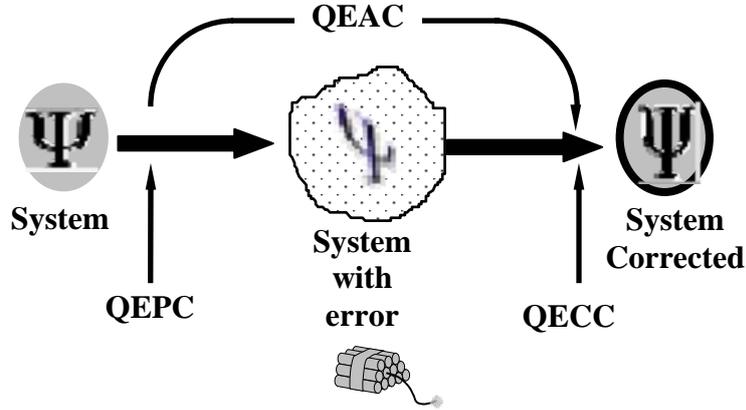

Fig. 5. Scheme of different strategies for error control.

Next we review each one of the strategies, placing special emphasis on the well developed QECC, although because the quantum circuits implementing them are expensive, they are giving way to other strategies which avoid errors.

## 8. Quantum error preventing codes (QEPC).

These are codes preventing the appearance of errors, although if they do take place, these codes are incapable of correcting them. They are based on the quantum Zeno effect: measuring repeatedly on a system, this continuously collapses freezing its evolution and avoiding the errors [11]. The use of this effect to prevent errors was suggested initially by Zurek [12].

Let us consider a system described by the initial state vector $|\phi(0)\rangle$, representing a quantum register of length n. Suppose the system evolves unitarily under the Hamiltonian $\hat{H} = \hat{H}_0 + \hat{H}_e$ (since there is no danger of confusion, we use the same notation as for a Hadamard rotation), where $\hat{H}_0$ describes a perfect evolution and $\hat{H}_e$ represents an error. Under these conditions, the state vector after a certain time δt, can be expressed as (with h/2π = 1):

$$|\phi(\delta t)\rangle = e^{-i\hat{H}\delta t}|\phi(0)\rangle = a(\delta t)|\phi(0)\rangle + b(\delta t)|\psi(0)\rangle \qquad (17)$$

where $|\psi(0)\rangle$ is an state orthogonal to $|\phi(0)\rangle$. After a time δt, the probability of obtaining the state $|\phi(0)\rangle$ when measuring on $|\phi(\delta t)\rangle$ is $|a(\delta t)|^2$ and its value can be expressed as $\langle\phi(0)|\exp(-i\hat{H}\delta t)|\phi(0)\rangle$. The probability for short times δt is:

$$|a(\delta t)|^2 \approx 1 - \langle(\hat{H} - \langle\hat{H}\rangle)^2\rangle \delta t^2 = 1 - (\Delta E)^2 \delta t^2 \qquad (18)$$

The probability that we project the state |ϕ(δt)⟩ on the subspace generated by {|ψ(0)⟩} (outside the subspace of interest generated by |ϕ(0)⟩) behaves like $O(\delta t^2)$. Sufficiently frequent measurements make the error probability as small as it is desired. This strategy is used in the stabilisation by symmetrisation method that could be considered as an extension of the majority voted method to the quantum scale. Next we consider the formalism introduced in [13].

Let us suppose that each computation time step has the probability of producing a correct result (1-η) (with η constant), after N steps, the probability of success is $(1-\eta)^N \sim \exp(-\eta N)$, decreasing exponentially with N. If we have a stabilisation method that diminishes the error by a factor 1/R per step, after N time steps, the probability of success will be exp(-ηN/R) which can be within a (1-δ) value, choosing R = ηN/-log(1-δ), having a polynomial dependence on N. Therefore an exponential error growth (such as it appears in the decoherence) can become stabilised by means of a method that reduces the error 1/R in each step. In this formalism, R is the redundancy introduced.

The application of this stabilisation method is as follows. If we carry out the same computation in R copies of our quantum computer, they work independently and without errors, the total state of the R computers will be the tensor product:

$$|\Psi(t)\rangle = |\phi(t)\rangle_{(1)} \otimes ... \otimes |\phi(t)\rangle_{(R)} \qquad (19)$$

where all |ϕ(t)⟩$_{(i)}$ represents the same state, introducing a certain type of quantum redundancy. This state, in which there is no error, belongs to a *symmetrical* subspace of whole Hilbert space $\mathcal{H}^{\otimes R}$. An error in a computation (or in all of them), would imply different vectors, so:

$$|\Psi(t)_e\rangle = |\phi(t)_1\rangle \otimes ... \otimes |\phi(t)_R\rangle > \qquad (20)$$

Defining a *symmetrical subspace* $\mathcal{H}_{SIM} \subset \mathcal{H}^{\otimes R}$ as the smaller subspace of $\mathcal{H}^{\otimes R}$ containing the vectors of the form:

$$\bigotimes_{i=1}^{R} |\chi\rangle_{(i)} \qquad (21)$$

projecting the noisy |Ψ(t)$_e$⟩ state into $\mathcal{H}_{SIM}$ would eliminate some of its errors.

In summary, the stabilisation method eliminates the possible errors projecting a state of R copies of our computer on the $\mathcal{H}_{SIM}$ subspace. The advantage of this process is that the dimension of $\mathcal{H}^{\otimes R}$ is $2^R$, whereas the one of $\mathcal{H}_{SIM}$ is R+1, if the dimension of $\mathcal{H}$ is 2. The $\mathcal{H}_{SIM}$ subspace has a dimension exponentially smaller than $\mathcal{H}^{\otimes R}$. Nevertheless, not all the errors are eliminated, since in $\mathcal{H}_{SIM}$ there are more vectors than those of the form |ϕ⟩⊗…⊗|ϕ⟩. In spite of that, it can be demonstrated that the error decreases by a factor R in each symmetrisation.

**9. Quantum error avoiding codes (QEAC).**

These are strategies that encode the information in states of certain subspaces that do not undergo decoherence, therefore they do not need to detect errors. These methods are useful with certain types of noise having some symmetry.

The idea arose in a work of Palma [14] and were called *avoiding codes*, later on to be called *decoherence free subspaces (DFS)* [15]. A simple model will clarify the main idea.

Let us suppose that single qubits undergo a decoherence introducing a random phase angle ϕ independent of the system space coordinates:

$$|0\rangle \rightarrow |0\rangle \quad \text{and} \quad |1\rangle \rightarrow e^{i\phi}|1\rangle \quad (22)$$

A qubit |q> = a|0> + b|1> put under this noise, suffers a rapid loss of coherence. The decoherence effect on a subspace of dimension 4, made up of two qubits is:

$$|00\rangle \rightarrow |00\rangle \quad |01\rangle \rightarrow e^{i\phi}|01\rangle$$
$$|10\rangle \rightarrow e^{i\phi}|10\rangle \quad |11\rangle \rightarrow e^{i2\phi}|11\rangle \quad (23)$$

Since the states |01> and |10> acquires the same phase, if we use the encoding $|0_E\rangle$ = |01> and $|1_E\rangle$ = |10>, a general qubit encoded as $|q_E\rangle$ = a$|0_E\rangle$ + b$|1_E\rangle$ evolves under the noise until the state $e^{i\phi}$ {a$|0_E\rangle$ + b$|1_E\rangle$}. The phase appearing has no importance and the subspace generated by {|01>, |10>}, is a decoherence free subspace.

The fact that the phase ϕ does not depend on space coordinates causes the decoherence to be invariant under qubit permutations. The recognition of such types of symmetries is what allows the introduction the decoherence free subspaces in which the system evolution is purely unitary.

## 10. Quantum error correcting codes (QECC)

A QECC can be defined as a pair $Q(\hat{E},\hat{R})$, made up of an encoding operation $\hat{E}$ and a recovery method $\hat{R}$. These are methods capable of detecting and correcting errors. Despite the impossibility of introducing redundancy such as in the classic codes, it is feasible to disperse the quantum information embodied in the qubit, allowing its recovery after undergoing certain types of errors. Given a qubit |q> = a|0> + b|1>, its encoding is an application $\hat{E}: \mathcal{H}^{\otimes k} \rightarrow \mathcal{H}^{\otimes n}$ from the Hilbert subspace of dimension k to a Hilbert space of a greater dimension n. The simplest case is to encode a single qubit (k=1) whereas n is the number of qubits in the code states (registers). Formally, to maintain the number of qubits in the application, (n-1) initial qubits |0> are introduced, and the qubit |q> can be encoded as:

$$\hat{E}\{(a|0\rangle + b|1\rangle) \otimes |0^{\otimes(n-1)}\rangle\} = |q_E\rangle = a|0_E\rangle + b|1_E\rangle \quad (24)$$

where $\hat{E}$ is the encoding operation and the qubits $|0_E\rangle$ and $|1_E\rangle$ are called *encoded*. The application only chooses an encoding subspace or *quantum code* $Q \subset \mathcal{H}^{\otimes n}$ of dimension two. So for the encoding is useful it must fulfil two conditions:

a) The error subspaces must be *distinguishable*.
   To identify the errors they must transform the encoded states of Q, to states of *mutually orthogonal subspaces* in $\mathcal{H}^{\otimes n}$.

b) *Maintaining the coherence.*
   The correction process must conserve the qubit coherence. Inside each orthogonal subspace, the total state must be the tensor product of the qubit and the environment state. This behaviour allows the erroneous qubit to be recovered by means of a measurement that projects the total state in one of those subspaces (see equation 13).

After the measurement, the qubit is uncoupled from the environment, and once the subspace on which we have projected is detected, we will be able to correct the error.

**10.1 Quasi-classic error correcting codes**

The simplest case of error correction consists of considering only bit-flip errors as in the classic case. Bit-flips attack the qubit $|q\rangle = a|0\rangle + b|1\rangle$ transforming it into $a|1\rangle + b|0\rangle$. We must be able to detect the error *without measuring destructively the qubit*, otherwise we would destroy its coherence. Next we review the fundamental steps of the whole process.

10.1.1 Error model

In addition to the aforementioned noise characteristics, we assume a symmetrical binary channel with an $\varepsilon$ (<0.5) error probability per qubit and time step. The purpose is to improve this level of error by means of an encoding and correction.

10.1.2 Encoding

Our starting point could be a classic binary repetition code [3,1,3] identifying each bit as a qubit. The encoding $\hat{E}$ will be:

$$|0\rangle \to |0_E\rangle = |000\rangle \quad \text{and} \quad |1\rangle \to |1_E\rangle = |111\rangle \quad (25)$$

A general qubit $|q\rangle$ is encoded as $\hat{E}(|q\rangle|00\rangle) = |q_E\rangle = a|0_E\rangle + b|1_E\rangle = a|000\rangle + b|111\rangle$. The information contained in the single qubit has been dispersed between three qubits, embedding the qubit into a two dimensional subspace (generated by $\{|000\rangle, |111\rangle\}$) of the $2^3=8$ dimensional Hilbert space, $\mathcal{H}^{\otimes 3}$. The set of correctable errors ($C_Q$) is made up of tensor products involving three factors including the identity (which is not an error itself) and a bit-flip error, represented by the $\hat{X}$ Pauli operator:

$$C_Q = \left\{ \hat{I} \otimes \hat{I} \otimes \hat{I} \equiv \hat{X}_{000}, \hat{I} \otimes \hat{I} \otimes \hat{X} \equiv \hat{X}_{001}, \hat{I} \otimes \hat{X} \otimes \hat{I} \equiv \hat{X}_{010}, \hat{X} \otimes \hat{I} \otimes \hat{I} \equiv \hat{X}_{100} \right\} \quad (26)$$

The operator subscript indicates the affected qubit. The code cannot correct other errors as we will see later. Notice that we could have chosen another base for the subspace or quantum code, for example $\{|000\rangle \pm |111\rangle\}$, but its correction capability is the same as the previous one, and both are equivalent codes.

10.1.3 Decoherence process

Sending a qubit $|q_E(0)\rangle = (a|0_E\rangle + b|1_E\rangle)$ through a depolarising noisy channel, produces an entanglement between the qubit and its environment:

$$|\Psi(t)\rangle = \begin{cases} (1-\varepsilon)^{3/2}|e_I\rangle\hat{I} + \\ + (1-\varepsilon)\sqrt{\varepsilon}\left[|e_X^{(1)}\rangle\hat{X}_{100} + |e_X^{(2)}\rangle\hat{X}_{010} + |e_X^{(3)}\rangle\hat{X}_{001}\right] + \\ + \varepsilon\sqrt{1-\varepsilon}\left[|e_X^{(12)}\rangle\hat{X}_{110} + |e_X^{(13)}\rangle\hat{X}_{101} + |e_X^{(23)}\rangle\hat{X}_{011}\right] + \\ + \varepsilon^{3/2}|e_X^{(123)}\rangle\hat{X}_{111} \end{cases} |q_E(0)\rangle \quad (27)$$

A particular case of the process would imply a single term describing a unitary error. If we can correct decoherence, we will be able to do it with the unitary errors.

10.1.4 Error detection.

The emitter sends the qubit $|q_E(0)\rangle$ through the noisy channel. The qubit-environment entanglement causes the receiver to detect the $|\Psi(t)\rangle$ state involving a linear combination of all possible bit-flip errors, each one with a certain coefficient related to its probability. To detect the error, the receiver would have to measure some of the qubits, but in doing so, it would collapse the state losing the information about the qubit coefficients (*destructive measurement*). We need another form to measure the qubits indirectly to *maintain the coherence*.

Instead of measuring destructively the $|\Psi(t)\rangle$ state, we can make a *collective measurement* that will allow the error syndrome to be obtained without acquiring knowledge about the qubit coefficients. A set of two CNOT gates (CNOT(1;2), CNOT(1;3)) could be used to translate the error syndrome to the last two qubits. After measuring them, the syndrome would permit us to recover the correct encoded qubit applying the appropriate $\hat{X}$ gates. Unfortunately this method has some drawbacks: it eliminates the encoding (i.e. the qubit protection) after the measure and worse, it will not be appropriate for fault-tolerant error correction (see section 10.6). In spite of the previous syndrome extraction, the receiver introduces two additional qubits or *ancilla*, in the initial state $|00_a\rangle$, preparing the state $|\Psi(t)\rangle|00_a\rangle$. In this code, the collective measurements consist of comparing the logical values of two pairs of qubits: the first and second and the first and third. The results are introduced in the ancilla qubits. In the collective measurement we are not interested in finding out the definite values of the qubits, only whether they are equal or different. The process is analogous to the classic case of the error syndrome measurement according to the parity check matrix:

$$H_C = \begin{pmatrix} 1 & 1 & 0 \\ 1 & 0 & 1 \end{pmatrix} \quad (28)$$

Altogether the process of syndrome extraction consists of an interaction $\hat{S}$ that permits the following operation to be carried out:

$$\hat{S}\left\{\hat{A}_i|q_E(0)\rangle \otimes |e_i\rangle \otimes |00_a\rangle\right\} = \hat{A}_i|q_E(0)\rangle \otimes |e_i\rangle \otimes |S_i\rangle \quad (29)$$

where the ancilla state $|S_i\rangle$ contains the syndrome information of the error and does not depend on the qubit state, but only on the error. When $\hat{S}$ is applied to the entangled state of the {qubit + environment} system (equation 27), we obtain:

$$|\Psi(t)\rangle|00_a\rangle \xrightarrow{\text{Collective Measurement}}$$

$$\begin{cases} (1-\varepsilon)^{3/2}|e_I\rangle\hat{I}\,|00_a\rangle + \\ + (1-\varepsilon)\sqrt{\varepsilon}\left[|e_X^{(1)}\rangle\hat{X}_{100}|11_a\rangle + |e_X^{(2)}\rangle\hat{X}_{010}|10_a\rangle + |e_X^{(3)}\rangle\hat{X}_{001}|01_a\rangle\right] + \\ + \varepsilon\sqrt{1-\varepsilon}\left[|e_X^{(12)}\rangle\hat{X}_{110}|01_a\rangle + |e_X^{(13)}\rangle\hat{X}_{101}|10_a\rangle + |e_X^{(23)}\rangle\hat{X}_{011}|11_a\rangle\right] + \\ + \varepsilon^{3/2}|e_X^{(123)}\rangle\hat{X}_{111}|00_a\rangle \end{cases} |q_E(0)\rangle \quad (30)$$

10.1.5 Syndrome extraction

The receiver measures the ancilla destructively on the computation basis $\{|0\rangle, |1\rangle\}$, collapsing the total state and obtaining two classic bits corresponding to the error syndrome.

Since the codewords have an *equal syndrome for the same error*, the measurement maintains coherence. Note that the four ancilla states identifying the error are orthogonal. This way of measuring solves the aforementioned third problem in the quantum error correction (section 4).

10.1.6 Error correction

Once the syndrome is measured, we correct the qubit state by applying the inverse unitary transformation: the identity $\hat{I}$ or a transformation $\sigma_X(i) \equiv \hat{X}_{(i)}$ is applied to the i qubit. For the previous repetition code, the collective measurement provides one syndrome for two different errors (implying that the code does not correct all the errors), nevertheless its probability (given by the square of the coefficient) is different, corresponding the applied correction to the error with highest probability. The code only corrects one bit-flip error, since the orthogonal states of the code, are transformed by the action of these errors into orthogonal states to each other and to the code itself. If this condition is fulfilled, the code is called *non-degenerate*.

With this code the correctable errors $C_Q$, transform the codewords $|0_E\rangle$, $|1_E\rangle \in Q \subset \mathcal{H}^{\otimes 3}$ (initially orthogonal), into orthogonal codewords to each other, as well as having the same syndrome if they come from the same error. Figure 6 shows how the three errors that can be corrected produce three orthogonal subspaces, each one with different syndrome.

The general error correction conditions are:

$$\forall \hat{A}_i \in C_Q \text{ and } \forall |u\rangle, |v\rangle \in Q \subset \mathcal{H}^{\otimes n} \Rightarrow \langle u|\hat{A}_i^+ \hat{A}_j|v\rangle = \delta_{ij}\delta_{uv} \qquad (31)$$

and for the present code are $\langle 1_E|\hat{X}_i^+ \hat{X}_j|1_E\rangle = \delta_{ij}$ and $\langle 0_E|\hat{X}_i^+ \hat{X}_j|1_E\rangle = 0$. The code cannot correct any two errors affecting the encoded qubits.

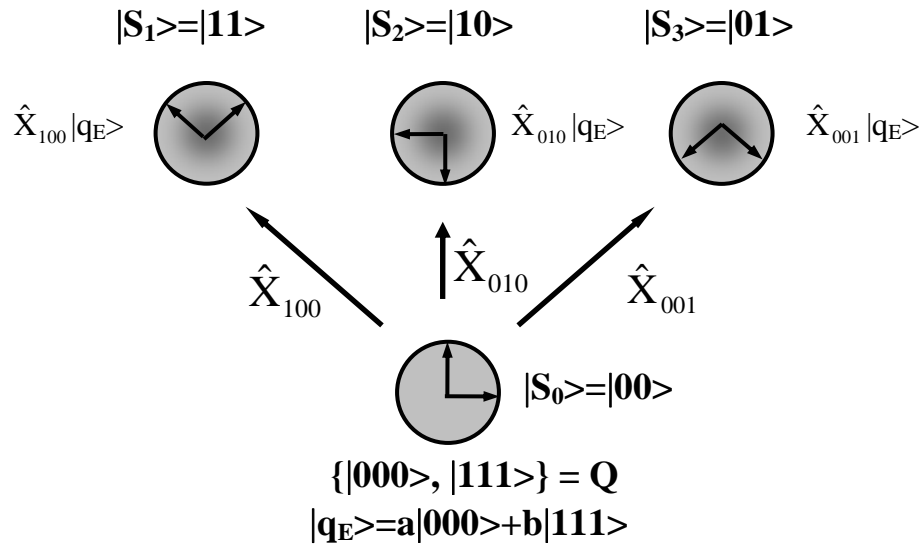

Fig. 6. Action of the bit-flip error operators on the Q subspace. The three error operators transform Q into mutually orthogonal subspaces.

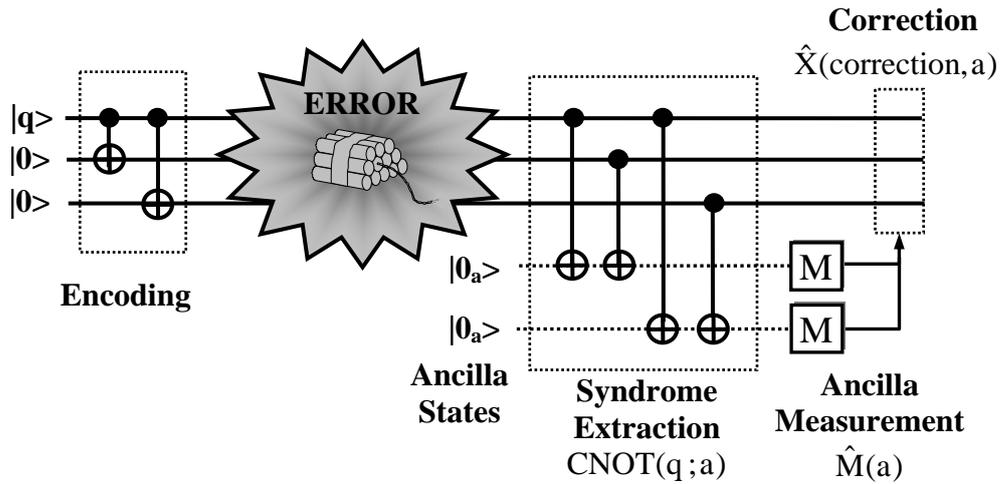

Fig. 7. Quantum circuit that implements the [[3,1,3]] encoding, syndrome extraction and qubit correction. Time flows from the left to the right.

   Finally we must bring back the ancilla (whose state contains the error syndrome) to its initial state $|00_a\rangle$ so as to be able to use it again. A cheaper possibility is to reject it and synthesise a new one. For the quantum codes a distance analogous to the classic case [16] (see section 2) can be defined. If the distance is $d \geq 2t+1$, the code is able to correct t errors in any one of the positions within the quantum register. Defining the weight of an error operator as the number of operators different from the identity in their tensor product, the value of t agrees with the weight of the error operators that the code can correct. In the present case, the quantum code has a distance 3 (with respect to the bit-flip errors), since it can correct one error, so Q is a code [[3,1,3]]. If we increase the number of qubits in this code, the distance increases. For example, a repetition code generated by $\{|00000\rangle, |11111\rangle\}$ has a distance 5 and corrects bit-flip errors of weight two.

10.1.7 Quantum circuit

   The syndrome is extracted by means of a set of CNOT gates between the qubit q (control) and the ancilla a (target). We can represent the syndrome extraction as the operator $\hat{S} = \text{CNOT}(q;a)$, constructed by means of the parity check matrix of the code $H_C$ (equation 28). The 1's in each row indicate the position of the control q-qubits, whereas target qubits are those of the a-ancilla. The circuit implementing the total process appears in figure 7.

   The recovery operator could be written as $\hat{R} = \hat{X}(\text{correction},a)\hat{M}(a)\hat{S}$ where $\hat{X}(\text{correction},a)$ represents the application of NOT gates depending on the syndrome contained in ancilla a and $\hat{M}(a)$ is an operator who describes the ancilla measurement. The former circuit is not unitary since it involves measurements. Although it is possible to construct a unitary circuit for the correction, the use of measurements has certain advantages when the tolerance to failures is taken into account.

   Note that the loss of information comes from the entanglement between the encoded qubit (quantum register) and its environment. Paradoxically, it is the entanglement between the ancilla and the register that allows us to recover the state if errors have taken place.

**10.2 Fidelity**

In the process of detection and error correction there is a probability that two or more different errors in qubits appear simultaneously. In order to measure the code capability to correct errors, *fidelity* can be used. This is defined as the minimum probability of obtaining the desired state of the system after a certain process has been carried out. In the present case the desired state is |q$_E$(0)> whereas the final state is a mixed one arising after measuring |Ψ(t)>. The probability that measuring |Ψ(t)> will collapse in the same initial state is <q$_E$(0)|tr$_e$ {|Ψ(t)><Ψ(t)|}q$_E$(0)>. The 'tr$_e$' means the partial trace over the environment states. As this value depends on the qubit coefficients a and b, we will choose the magnitude which characterises how good the process is, as the minimum value with respect to all the possible states:

$$\text{Fidelity} = \underset{\forall |q_E(0)\rangle}{\text{Min}} \left\{ \langle q_E(0) | tr_e \left[ |\Psi(t)\rangle\langle\Psi(t)| \right] | q_E(0) \rangle \right\} \qquad (32)$$

Fidelity does not depend on the considered initial state, but only on the particular process, through |Ψ(t)>. The main objective of error correction is to maximise the fidelity.

Considering only bit-flip errors, if we sent a qubit |q(0)> without encoding (nor using error correction) the fidelity would be:

$$F_{WE} = \underset{\forall |q(0)\rangle}{\text{Min}} \left\{ (1-\varepsilon) + \varepsilon \left| \langle q(0) | \hat{X} | q(0) \rangle \right|^2 \right\} = 1 - \varepsilon \qquad (33)$$

Since the second term is positive and its minimum value corresponds to the case |q(0)> = |0> with zero value, the fidelity behaves as $F_{WE} \sim 1-O(\varepsilon)$.

Let us assume now that we encode the qubit |q(0)> with a quantum code Q = [[3,1,3]], that corrects one bit-flip error in any one of the three qubits in the register |q$_E$(0)>. Supposing that the correction process is error free, all the errors affecting one qubit can be eliminated, which is reflected in the term $3\varepsilon(1-\varepsilon)^2$ of the (encoded) fidelity:

$$F_E = \underset{\forall |q_E(0)\rangle}{\text{Min}} \left\{ (1-\varepsilon)^3 + 3\varepsilon(1-\varepsilon)^2 + \text{positive terms} \right\} \qquad (34)$$

The positive terms are zero in the least favourable situation, so the fidelity in the encoded case is $F_E = (1-\varepsilon)^3 + 3\varepsilon(1-\varepsilon)^2 \sim 1 - O(\varepsilon^2)$, disappearing the linear term in ε. So that $F_E > F_{WE}$ must be fulfilled, ε < 0.5 is required (as in the classic case).

**10.3 Error correcting codes for phase-flip errors.**

Phase-flip errors are typically quantum, although their correction is related to the bit-flip errors. They arise when the entanglement of the system with its environment, gives rise to a phase decoherence. The general noise characteristics considered are the same as those of the previous case.

In order to look for the appropriate encoding, we see that there is a close relationship between the bit-flip errors and those of phase-flip, through the form of the operators who produce them. The phase-flip errors can be represented by $\hat{Z}$ operators, but $\hat{Z} = \hat{H}\hat{X}\hat{H}$, where $\hat{H}$ is a Hadamard rotation. We use as codewords of the new $Q_f$ code $\hat{H}^{\otimes n}\{|0_E\rangle, |1_E\rangle\}$, where |0$_E$> and |1$_E$> are codewords of a code $Q_b$ correcting single bit-flip errors (and therefore with minimum distance 3). Encoding the qubit |q> provides |q$_E$> = $a\hat{H}^{\otimes n}|0_E\rangle + b\hat{H}^{\otimes n}|1_E\rangle$. If the

channel introduces a phase-flip error $\hat{Z}_e$ in the qubit $|q_E\rangle$, we will have $\hat{Z}_e|q_E\rangle = \hat{Z}_e\{a\hat{H}^{\otimes n}|0_E\rangle + b\hat{H}^{\otimes n}|1_E\rangle\}$ and applying the recovery operator

$$\hat{R} = \{\hat{H}^{\otimes n}\, \hat{X}(\text{correction}, a)\, \hat{M}(a)\, \hat{S}\, \hat{H}^{\otimes n}\} =$$
$$= \{\hat{H}^{\otimes n}\, \hat{X}(\text{correction}, a)\, \hat{M}(a)\, \text{CNOT}(q;a)\, \hat{H}^{\otimes n}\} \qquad (35)$$

we obtain:

$$\hat{R}\{\hat{Z}_e|q_E\rangle\} =$$

$$= \{\hat{H}^{\otimes n}\, \hat{X}(\text{correction}, a)\, \hat{M}(a)\, \text{CNOT}(q;a)\, \hat{H}^{\otimes n}\}\{\hat{Z}_e(a\hat{H}^{\otimes n}|0_E\rangle + b\hat{H}^{\otimes n}|1_E\rangle)\}|00_a\rangle = \qquad (36)$$

$$= \hat{H}^{\otimes n}\, \hat{X}(\text{correction}, a)\, \hat{M}(a)\, \text{CNOT}(q;a)\{a\hat{X}_e|0_E\rangle + b\hat{X}_e|1_E\rangle\}|00_a\rangle =$$

$$= \hat{H}^{\otimes n}\hat{X}(\text{correction}, a = S_e)\{a\hat{X}_e|0_E\rangle + b\hat{X}_e|1_E\rangle\}|S_e\rangle = \{a\hat{H}^{\otimes n}|0_E\rangle + b\hat{H}^{\otimes n}|1_E\rangle\}|S_e\rangle$$

The CNOT(q;a) operation on the codewords $|0_E\rangle$ and $|1_E\rangle$ of $Q_b$, copy the bit-flip error information of the qubit q (control) onto the ancilla a (target), in accordance with the parity check matrix. The operator $\hat{M}(a)$ represents the ancilla measurement (whose result is the error syndrome $|S_e\rangle$) and the $\hat{X}(\text{correction}, a = S_e)$ represent the correction depending on the ancilla measurement result. Finally, the encoded qubit is restored to the original encoded base $\hat{H}^{\otimes n}\{|0_E\rangle, |1_E\rangle\}$.

If we take $Q_b = \{|000\rangle = |0_E\rangle, |111\rangle = |1_E\rangle\}$, the new codewords of $Q_f$ are:

$$\hat{H}^{\otimes 3}|000\rangle = |0_E^f\rangle = \frac{1}{\sqrt{2}}(|0\rangle + |1\rangle)\frac{1}{\sqrt{2}}(|0\rangle + |1\rangle)\frac{1}{\sqrt{2}}(|0\rangle + |1\rangle) =$$
$$= \frac{1}{\sqrt{8}}\{|000\rangle + |001\rangle + |010\rangle + |100\rangle + |011\rangle + |101\rangle + |110\rangle + |111\rangle\} \qquad (37a)$$

$$\hat{H}^{\otimes 3}|111\rangle = |1_E^f\rangle = \frac{1}{\sqrt{2}}(|0\rangle - |1\rangle)\frac{1}{\sqrt{2}}(|0\rangle - |1\rangle)\frac{1}{\sqrt{2}}(|0\rangle - |1\rangle) =$$
$$= \frac{1}{\sqrt{8}}\{|000\rangle - |001\rangle - |010\rangle - |100\rangle + |011\rangle + |101\rangle + |110\rangle - |111\rangle\} \qquad (37b)$$

With this code $Q_f = \{\hat{H}^{\otimes 3}|000\rangle, \hat{H}^{\otimes 3}|111\rangle\}$, the qubit $|q\rangle = a|0\rangle + b|1\rangle$, is encoded as $|q_E^f\rangle = a|0_E^f\rangle + b|1_E^f\rangle$. The two codewords of $Q_f$ are also orthonormal.

Sending the qubit $|q_E^f\rangle$ through a noisy channel that introduces phase-flip errors, an entanglement with its environment occurs, similarly to that established in the previous code. The difference is replacing the operators $\hat{X}_{ijk}$ for $\hat{Z}_{ijk}$ and their correlated environment states in equation (27).

The set of correctable errors of $Q_f$ is:

$$C_{Q_f} = \left\{ \hat{I} \otimes \hat{I} \otimes \hat{I} \equiv \hat{Z}_{000}, \hat{I} \otimes \hat{I} \otimes \hat{Z} \equiv \hat{Z}_{001}, \hat{I} \otimes \hat{Z} \otimes \hat{I} \equiv \hat{Z}_{010}, \hat{Z} \otimes \hat{I} \otimes \hat{I} \equiv \hat{Z}_{100} \right\} \quad (38)$$

and the code for the phase-flip errors is $Q_f$ = [[3,1,3]]. Just like the $Q_b$ = Q code, there are errors that cannot be corrected, but their weight is bigger than those can be corrected and the encoded fidelity behaves like 1-O($\varepsilon^2$).

The syndrome measurement circuit and qubit correction implementing $\hat{S}$ is analogous to the one in the previous case (figure 7), with the difference that the encoding is carried out in the base $\{|0_E^f>, |1_E^f>\}$ and three Hadamard gates must appear just before and after the error correction.

**10.4 Phase and bit-flip error correcting codes.**

The correction power of the previous codes is limited. The code $Q_b$ = [[3,1,3]] uses qubit redundancy to correct a single bit-flip error; the $Q_f$ = [[3,1,3]] uses sign redundancy to correct a single phase-flip error. Nevertheless we must find a *single* quantum code capable of correcting *both* types of errors. Historically it was Shor [17] who in 1995 introduced the first code that did what for some time was thought impossible: correcting quantum errors. The encoding was:

$$\hat{E}\{|0\rangle \otimes |0^{\otimes 8}\rangle\} = |0_E\rangle = \frac{1}{\sqrt{2}} (|000\rangle + |111\rangle)(|000\rangle + |111\rangle)(|000\rangle + |111\rangle)$$
$$\hat{E}\{|1\rangle \otimes |0^{\otimes 8}\rangle\} = |1_E\rangle = \frac{1}{\sqrt{2}} (|000\rangle - |111\rangle)(|000\rangle - |111\rangle)(|000\rangle - |111\rangle)$$
(39)

So a qubit |q> = a|0> + b|1> is encoded into $|q_E>$ = $a|0_E>$ + $b|1_E>$. If there appears a bit-flip error in some set of three qubits, it is possible to detect and correct it, by means of an analogous method used with Q. If a phase-flip error happens in one of these three sets, and we have some strategy to compare the sets, we will be able to detect and correct them. Note that in Shor's code some errors as $\hat{Z}_{110}$, $\hat{Z}_{101}$ or $\hat{Z}_{011}$, even though they do not produce orthogonal states, they are equivalent (equal) and correctable. These codes are called *degenerated*.

Almost simultaneously Steane (1996) introduced a method for transforming certain types of classic codes into quantum ones. The idea that guided him was that bit-flip errors could be corrected with a code of classic type and the phase-flip errors were equivalent to bit-flips if a Hadamard rotation were previously made. When rotating the codewords, it had to make sure that they did not leave some code of suitable distance.

Steane encoded two qubits |0> and |1> starting with a classic Hamming code C = [7,4,3] containing its dual $C^\perp$ = [7,3,4] (even subcode, since it contains only the codewords of even weight). The basis of the quantum code include two entangled states obtained from the classic codewords of each coset of C relative to $C^\perp$: $C^\perp \oplus (0000000) = C^\perp$ = {codewords of C with even weight }, and the $C^\perp \oplus (1111111)$ = {codewords of C with odd weight} (see figure 8). The quantum codewords are:

$$|0_E\rangle = |C^\perp\rangle = \frac{1}{\sqrt{8}}\left\{\begin{array}{l}|0000000\rangle + |0001111\rangle + |0110011\rangle + |0111100\rangle \\ |1010101\rangle + |1011010\rangle + |1100110\rangle + |1101001\rangle\end{array}\right\}$$
(40)

$$|1_E\rangle = |C^\perp \oplus (1111111)\rangle = \frac{1}{\sqrt{8}}\left\{\begin{array}{l}|1111111\rangle + |1110000\rangle + |1001100\rangle + |1000011\rangle \\ |0101010\rangle + |0100101\rangle + |0011001\rangle + |0010110\rangle\end{array}\right\}$$

The vector space generated by the (encoded) computation base F = {|0$_E$>, |1$_E$>} corresponds to a quantum code Q (analogous to Q$_b$) correcting one bit-flip. In addition to the F base, we can use other bases, for example the dual (encoded) base P = {$\hat{H}^{\otimes 7}$|0$_E$>, $\hat{H}^{\otimes 7}$|1$_E$>}:

$$\hat{H}^{\otimes 7}|0_E\rangle = \frac{1}{\sqrt{2}}\{|0_E\rangle + |1_E\rangle\} \qquad \hat{H}^{\otimes 7}|1_E\rangle = \frac{1}{\sqrt{2}}\{|0_E\rangle - |1_E\rangle\} \quad (41)$$

consisting of two entangled and orthonormal states involving codewords of the [7,4,3] classic code that can correct one bit-flip error.

10.4.1 Detection and error correction

Since the quantum encoding uses linear combinations of classic codewords (in C) of distance 3, it is possible to detect single bit-flip errors. The appearance of an $\hat{X}_e$ error (the error is applied to the qubits where the vector e ∈ GF(2)$^7$ has 1's), moves the codewords {|0$_E$>, |1$_E$>} towards $\hat{X}_e$ {|0$_E$>, |1$_E$>}, both in same coset of C, maintaining coherent superpositions. In order to measure the syndrome, an ancilla with three qubits (|000$_a$>) is used into which the syndrome is copied by means of CNOT gates placed according to the parity check matrix of C = [7,4,3]:

$$H_{[7,4,3]} = \begin{pmatrix} 1 & 0 & 1 & 0 & 1 & 0 & 1 \\ 0 & 1 & 1 & 0 & 0 & 1 & 1 \\ 0 & 0 & 0 & 1 & 1 & 1 & 1 \end{pmatrix} \quad (42)$$

The measurement of the ancilla qubits provides the syndrome bits, in accordance with which NOT gates are applied ($\hat{X}(\text{correction}, a)$) there where necessary. The bit-flip errors produce the effect $\hat{X}_e |q_E\rangle = |q_E \oplus e\rangle$, and the correction can be outlined as:

$$\hat{R}\{\hat{X}_e|q_E\rangle\} = \{\hat{X}(\text{correction}, a)\,\hat{M}(a)\,\text{CNOT}(q;a)\}\,|q_E \oplus e\rangle|000_a\rangle \rightarrow \quad (43)$$
$$\rightarrow \hat{X}(\text{correction}, a = S_e)\,|q_E \oplus e\rangle|S_e\rangle = |q_E\rangle|S_e\rangle$$

A phase-flip error transforms the qubit into $\hat{Z}_e|q_E\rangle$. Its detection involves a seven qubits Hadamard rotation. By virtue of the $\hat{H}\hat{Z}_e = \hat{X}_e\hat{H}$ condition, we transform phase-flip errors in base F into bit-flip errors in the P base. The relationship between both bases can be understood easily. A phase-flip error in the computation base F (|0> → |0> and |1> → -|1>), corresponds to a bit-flip error in the dual base P ($\hat{H}$|0> → $\hat{H}$|1> and $\hat{H}$|1> → $\hat{H}$|0>). As the

P base involves C codewords of distance 3, it is possible to make a correction for bit-flips using a three qubit ancilla state. We apply a set of CNOT(q;a) gates to obtain the error syndrome and correct with $\hat{X}(\text{correction}, a = S_e)$ gates. To conclude, we rotate back the qubit state to let it in the original computation base. The complete correction circuit is shown in figure 9.

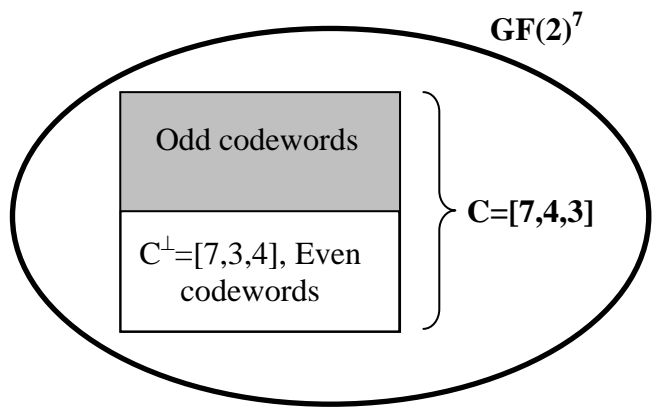

Fig. 8. Relationship between the $GF(2)^7 = \{0, 1\}^{\otimes 7}$ vector space and the subspaces conforming the [7,4,3] Hamming code and its dual.

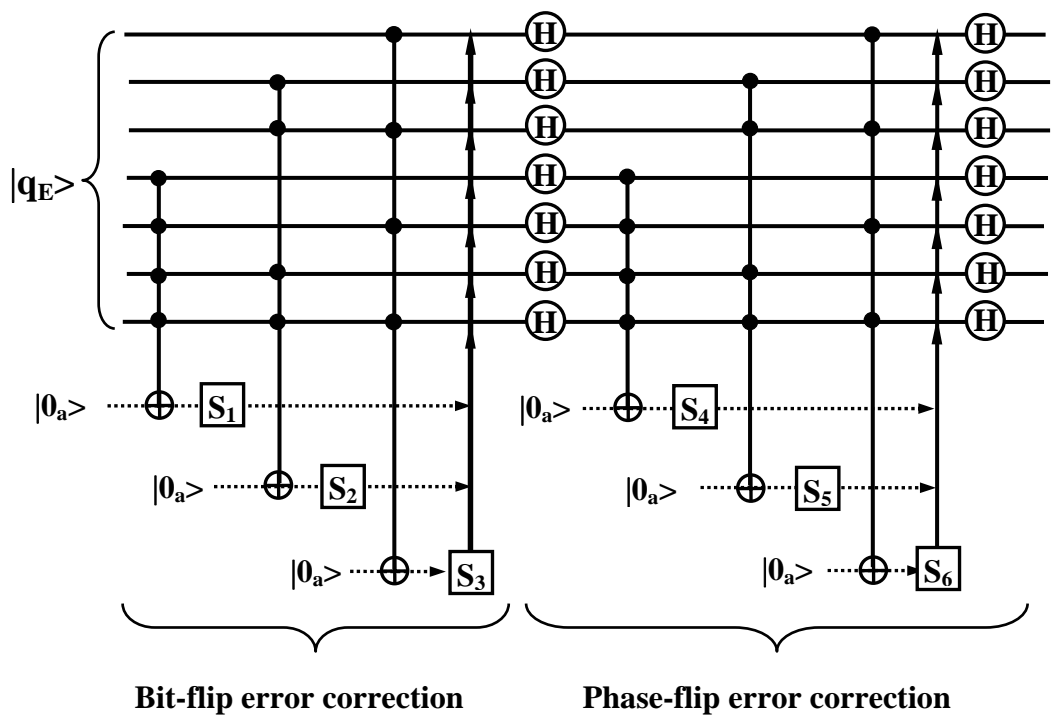

Fig. 9. Quantum circuit implementing the syndrome extraction and qubit correction when it is encoded ($|q_E\rangle$) by means of Steane's [[7,1,3]] quantum code. In order to measure the syndrome for the bit-flip and phase-flip errors, six ancillas in the initial state $|0_a\rangle$ are used.

The syndrome ($S_1$, $S_2$, $S_3$) describe bit-flip errors, whereas ($S_4$, $S_5$, $S_6$) correspond to the phase-flip errors. Correcting both, it is also done for the $\hat{Y}_e$ errors, because $\hat{Y}_e = \hat{Z}_e \hat{X}_e$. We conclude that the Steane code is [[7,1,3]] and corrects any $\hat{X}$, $\hat{Y}$ and $\hat{Z}$ error.

### 10.4.2 CSS codes

The construction method of the Steane [[7,1,3]] code, can become general to obtain other codes. We now describe a family of codes called CSS, whose design is based on the theory of classic linear codes. Discovered by Calderbank, Shor [18] and Steane [19], being the Steane's code a particular case, the method is founded on the theorem of the dual code.

*Theorem of the dual code*: By rotating Hadamard a quantum state obtained as the linear combination of all the codewords of a linear classic code C = [n,k,d], we get a state which is the linear combination of all the codewords of its dual $C^\perp$ (linear) code:

$$\hat{H}^{\otimes n} \frac{1}{\sqrt{2^k}} \sum_{i \in C} |i\rangle = \frac{1}{\sqrt{2^{n-k}}} \sum_{x \in C^\perp} |x\rangle \qquad (44)$$

The CSS construction is as follows. Consider two classic linear codes: $C_1 = [n,k_1,d_1]$ whose parity check matrix is $H_1[(n-k_1) \times n]$ and $C_2$ with parity check matrix $H_2[(n-k_2) \times n]$ is such that $C_2$(subcode) $\subseteq C_1$. Then $k_2 < k_1$ and the parity check matrix of $H_2$ contains $(n-k_1)$ rows of $H_1$ and some other $(k_1-k_2)$ linearly independent rows assuring $C_2 \subseteq C_1$. The subcode $C_2$ defines an equivalence relationship $\Re$ in $C_1$: $\forall u, v \in C_1$, $u \Re v \Leftrightarrow u-v \in C_2$, or which is the same $u \Re v \Leftrightarrow$ if $\exists w \in C_2 \mid u = v + w$. The equivalence classes are cosets of $C_1$ relative to $C_2$ (elements of the factor group $C_1/C_2$). The number of cosets is $2^{k_1}/2^{k_2} = 2^{k_1-k_2}$. Let us transform the classic codewords of coset $C_2 \oplus a$ ($a \in C_1$) into quantum states and construct an entangled state of the type:

$$|C_2 \oplus a\rangle = \frac{1}{\sqrt{2^{k_2}}} \sum_{i \in C_2} |i \oplus a\rangle \qquad (45)$$

The set of these states forms an orthonormal base of a subspace of $2^{(k_1-k_2)}$ dimension of the Hilbert space $\mathcal{H}^{\otimes n}$ (see figure 10). The states $|C_2 \oplus a\rangle$ are created by the linear combination of distance $d_2$ codewords of a $C_2$ code, so it will be capable of correcting $t_2 = \lfloor (d_2-1)/2 \rfloor$ bit-flip errors. In addition, as the syndrome of all the codewords depends solely on the error, the syndrome extraction will maintain the qubit coherence. In general we can provide the following

*Definition*: Given two linear classic codes $C_1 = [n,k_1,d_1]$ and $C_2 = [n,k_2,d_2]$ (its dual being $C_2^\perp = [n, n-k_2, d_2^\perp]$) so that $C_2$(subcode) $\subseteq C_1$, the subspace generated by the encoded base $\{|C_2 \oplus a\rangle, a \in C_1\}$ is a *quantum CSS code* $Q(C_1, C_2) = [[n,k_1-k_2,D]]$ of dimension $2^{(k_1-k_2)}$ and distance $D \geq \text{Min}\{d_1, d_2^\perp\}$.

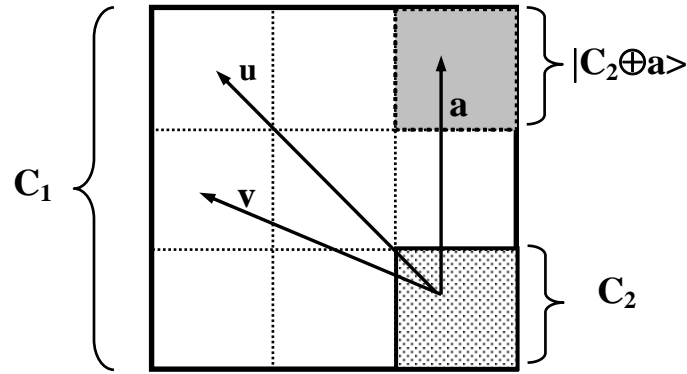

Fig. 10. Construction process of a CSS code from the code $C_2 \subseteq C_1$. Each box represents a coset with a different syndrome depending on the displacement vector

In order to construct quantum codes with this method it is sufficient to look for classic codes contained in its dual (or vice versa). Given a classic code C, if $C \subset C^\perp$ is fulfilled, it is called *weakly self-dual*. An example of weakly self-dual and self-dual linear binary codes (C = $C^\perp$) that cover a large interval of distances and code rate (k/n) is the family of Reed-Muller codes (RM) [20]. Starting with self-duals RM codes, quantum codes of dimension one can be constructed as [[n,0,d]]. From the [8,4,4] [32,16,8] and [128,64,16] RM codes, we obtain [[8,0,4]], [[32,0,8]] and [[128,0,16]] respectively. With the purpose of obtaining codes with dimension two, the self-dual RM codes can be punctured, whose dual code is an even subcode. Puncturing (deleting coordinates) the [8,4,4] we get [7,4,3] that contains the even subcode [7,3,4], providing the well-known [[7,1,3]] Steane quantum code. From the other weakly self-dual RM codes, the [[31,1,7]] and [[127,1,15]] are derived, correcting errors of weight 3 and 7 respectively. From RM codes of greater dimension as [64,42,8] (whose dual is [64,22,16]), other quantum codes can be obtained as [[64,20,8]].

10.4.3 Stabiliser codes [21]

Quantum codes are certain vector subspaces of $\mathcal{H}^{\otimes n}$. A way of specifying them is as the common eigenspaces of a set of commuting operators forming itself an abelian sub-group (called *stabiliser group* $S_Q$) of the Pauli group. The Pauli group $G_n$ is made up of the operators $\{\pm 1\} \times \{ \hat{A}_{\{i_1,i_2,...,i_n\}} = \hat{A}^1_{i_1} \otimes \hat{A}^2_{i_2} \otimes ... \otimes \hat{A}^n_{i_n} \}$.

In the case of the repetition code [[3,1,3]], we have a Hilbert space of dimension $2^3$. If we want to specify the code as a subspace of dimension 2, we can use the eigenspace common to two operators. For example, the common eigenspace of the set $\{\hat{Z}_{110}, \hat{Z}_{101}\}$ is the code Q = $\{|000>, |111>\} \equiv [[3,1,3]]$, that is where an encoded qubit resides when it does not have errors. The set can be transformed into a group $S_Q$ if the product of its operators is included. This $S_Q$ group is abelian and is called *stabilizer*, because its operators fix the codewords of the quantum code Q. Actually $S_Q$ is a subgroup of the $G_n /\{\pm\hat{I}\}$ factor group. The $\{\pm\hat{I}\}$ is the centralizer of $G_n$, so that we do not mind of the global operator phase and $S_Q$ is abelian. $S_Q$ can be specified completely by its generators $S_Q = \langle \hat{Z}_{110}, \hat{Z}_{101} \rangle$ (the notation <...> is used to specify the group generators.).

If an encoded qubit |q_E> undergoes an error $\hat{X}_v$, its state becomes $\hat{X}_v|q_E\rangle$, and is fixed by $\hat{X}_v S_Q \hat{X}_v = \langle \hat{X}_v \hat{Z}_{110} \hat{X}_v, \hat{X}_v \hat{Z}_{101} \hat{X}_v \rangle$ because:

$$\hat{X}_v \hat{Z}_u \hat{X}_v (\hat{X}_v|q_E\rangle) = \hat{X}_v \hat{Z}_u |q_E\rangle = \hat{X}_v |q_E\rangle \quad (u = 110, 101) \quad (46)$$

and $\hat{X}_v|q_E\rangle \in \hat{X}_v Q$. The syndrome is determined by the existence of an operator in $S_Q$ anticommuting with the error operator $\hat{X}_v$. If $\hat{X}_v = \hat{X}_{100}$:

$$\{\hat{X}_{100}, \hat{Z}_u\} = \{\hat{X}_{100}, \hat{Z}_{100}\} \hat{Z}_{(100)\oplus u} = 0 \quad (u = 110, 101) \quad (47)$$

since $\hat{X}_{100}$ commute with $\hat{Z}_{010}$ and $\hat{Z}_{001}$:

$$\hat{Z}_{101}(\hat{X}_{100}|q_E\rangle) = \hat{Z}_{100}\hat{X}_{100}\hat{Z}_{001}|q_E\rangle = -\hat{X}_{100}\hat{Z}_{101}|q_E\rangle = -\hat{X}_{100}|q_E\rangle = (-1)^a \hat{X}_{100}|q_E\rangle \quad (48a)$$

$$\hat{Z}_{110}(\hat{X}_{100}|q_E\rangle) = -\hat{X}_{100}\hat{Z}_{110}|q_E\rangle = -\hat{X}_{100}|q_E\rangle = (-1)^b \hat{X}_{100}|q_E\rangle \quad (48b)$$

The syndrome of the $\hat{X}_{100}$ error is (a,b) = (1,1). An error operator anticommuting with an operator in $S_Q$, changes the eigenvalue of the state from +1 to –1. Figure 11 shows the single bit-flip error syndromes and their orthogonal subspaces.

In the case of the Steane code, the stabiliser is generated by 6 operators (obtained replacing the 1's in the rows of $H_{[7,4,3]}$, equation (42), by $\hat{X}$ or $\hat{Z}$ operators), whose common eigenspace with eigenvalue +1, makes up the code [[7,1,3]]. Shor's [[9,1,3]] code can be described by means of a stabiliser with 8 generators. CSS codes are stabilizers; nevertheless these last ones contain other codes that are not CSS. For example the *perfect* quantum code [[5,1,3]] [22] (saturates the quantum Hamming bound $2(1+3n) \leq 2^n$ for codes with d=3, [23], analogous to classic equation 5), is not a CSS code although it is a stabilizer one.

| \|000>, \|111><br>$\hat{I}Q$<br>(a,b)=(0,0) | \|001>, \|110><br>$\hat{X}_{001}Q$<br>(0,1) |
|---|---|
| \|010>, \|101><br>$\hat{X}_{010}Q$<br>(1,0) | \|100>, \|011><br>$\hat{X}_{100}Q$<br>(1,1) |

Fig. 11. Relationship of the different subspaces from the repetition code that corrects one bit-flip. The pairs in parenthesis indicate the error syndrome in each subspace. For the code, the syndrome is (a,b) = (0,0).

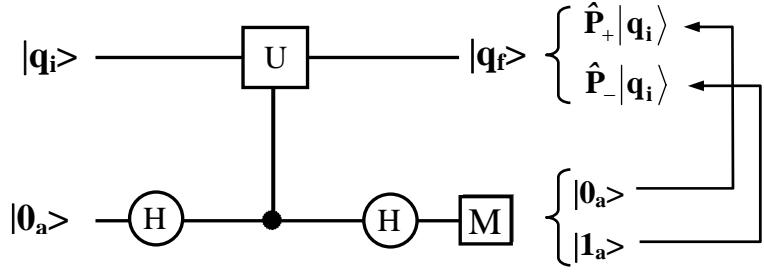

Fig. 12. Measurement circuit for the hermitian operator $\hat{U}$. The gate connecting both qubits is a control-U.

Given that the errors change the eigenvalue of the $S_Q$ generators, the correction circuit construction can be described in a more general way as the collective measurement of these operators. The measurement of operators is a fundamental element in error correction. The objective is to project the qubit state on an eigenstate of $S_Q$, at the same time as we keep an indicator from the eigenvalue in some quantum register. Let us suppose that we have a hermitian operator (such as an observable) and unitary (which can also represent a time evolution) $\hat{U}$, having the ±1 eigenvalues. In order to measure $\hat{U}$ we must make a projection of the qubit on one of its two eigenspaces. The circuit implementing the measurement appears in figure 12.

The initial state of the qubit is $|q_i\rangle$ and an ancilla in the $|0\rangle$ state is used. Its joint evolution is:

$$|q_i\rangle|0_a\rangle = \{a|0\rangle + b|1\rangle\}|0_a\rangle \xrightarrow{\hat{I}\otimes\hat{H}} \{a|0\rangle + b|1\rangle\}\frac{1}{\sqrt{2}}\{|0_a\rangle + |1_a\rangle\}$$

$$\xrightarrow{CU(2;1)} \frac{1}{\sqrt{2}}\{a|0\rangle|0_a\rangle + a(\hat{U}|0\rangle)|1_a\rangle + b|1\rangle|0_a\rangle + b(\hat{U}|1\rangle)|1_a\rangle\} \qquad (49)$$

$$\xrightarrow{\hat{I}\otimes\hat{H}} \frac{1}{2}\{[a|0\rangle + b|1\rangle + a\hat{U}|0\rangle + b\hat{U}|1\rangle]\otimes|0_a\rangle + [a|0\rangle + b|1\rangle - a\hat{U}|0\rangle - b\hat{U}|1\rangle]\otimes|1_a\rangle\}$$

The CU(2;1) means a control-U gate acting on the qubit 1 (target) depending on the qubit 2 value (control). The projectors on the eigenspaces with eigenvalues ±1 are $\hat{P}_\pm = (\hat{I}\pm\hat{U})/2$. Measuring the ancilla qubit of the previous evolution, if we obtain the state $|0_a\rangle$, we will have projected according to $\hat{P}_+$, and if the result is $|1_a\rangle$, the projection will correspond to $\hat{P}_-$. Note that the qubits used can be either non-encoded or encoded. Using this construction we obtain the circuit shown in figure 13. To determine the syndrome, the operators $\hat{Z}_{110}$ and $\hat{Z}_{101}$ are measured by means of two ancilla qubits initially in the state $|00_a\rangle$. Bearing in mind the equivalence $\hat{Z} = \hat{H}\hat{X}\hat{H}$ (figure 14), it is easy to get the circuit as it appears in figure 7.

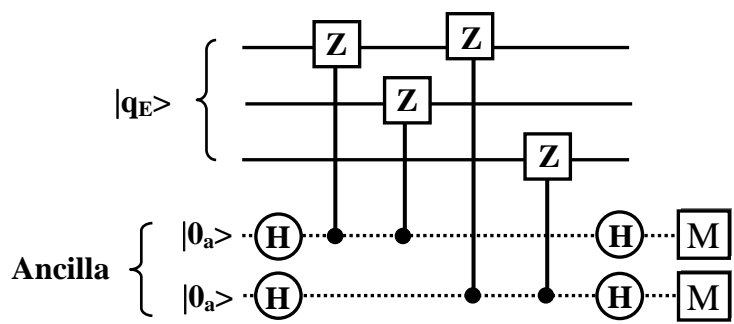

Fig. 13. Quantum circuit measuring the generators $\hat{Z}_{110}$ and $\hat{Z}_{101}$ of the [[3,1,3]] code. Gates M provide the error syndrome.

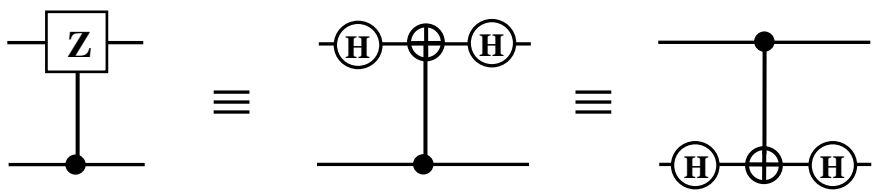

Fig. 14. Equivalence of circuits used in the syndrome extraction.

### 10.4.4 Codes on GF(4)

It is possible to relate the stabiliser quantum codes and the classic codes in GF(4) [24], establishing an isomorphism between the elements of the stabiliser and those of a subcode of $GF(4)^n$ that is selforthogonal with respect to certain symplectic product. This is the case in the [[5,1,3]] code that comes from a Hamming code in GF(4). This connection with the classic codes has allowed the well known constructions of these codes to be used to obtain a great amount of new quantum codes with a distance greater than 3, correcting more than one error. In the same way, stabilizer codes can be generalized to nonbinary alphabets over finite fields [25].

### 10.5 Quantum operation formalism applied to QECC

The fundamental pieces of quantum error correction are quantum states to be protected and noise. There are several ways to point out the theory [26]. Until now we have used state vectors or kets emphasizing the environment effect on the studied system as originating the errors. Experimentally it is not possible to know environment states, then a formalism based on the concept of *quantum operations* (or superoperators, see Knill and Laflamme in [3]) [27] will be more general and powerful to treat the evolution of open systems such as quantum computers.

In general, quantum states are described by means of the density operator $\hat{\rho}$ (or density matrix, if a basis set is chosen), and its time evolution by the quantum operation E defined as a map $\hat{\rho} \mapsto \mathsf{E}(\hat{\rho})$, an which has the following properties:

1.- E is a *convex-linear map* in the density operator set, fulfilling:

$$E\left(\sum_i p_i \hat{\rho}_i\right) = \sum_i p_i E(\hat{\rho}_i) \qquad (50)$$

{$p_i$} being the probability set for the { $\hat{\rho}_i$ } states.

2.- E is *completely positive map*. $E(\hat{O}_S)$ is more than a positive operator for any positive operator $\hat{O}_S$ of the system S. Consider all possible extensions T of S to the combined system TS, then E is completely positive in S if $(I_T \otimes E)(\hat{O}_{TS})$ is positive for any positive operator $\hat{O}_{TS}$ of TS.

3.- The value $0 \leq tr[E(\hat{\rho})] \leq 1$, is the probability that the process represented by E occur when $\hat{\rho}$ is its initial density operator.

Analogous to the one qubit evolution of equation 10, the evolution of a system such as an n-qubit register |$q_1,\ldots,q_n$> in contact with the environment (in the initial state |e>) can be expressed as:

$$\hat{U}\{|q_1,\ldots,q_n\rangle \otimes |e\rangle\} = \sum_i \left(|\mu(t)_i\rangle \otimes \hat{B}_i\right)|q_1,\ldots,q_n\rangle \qquad (51)$$

An orthonormal environment basis set {|$\mu_i$>} has been used. Now the operators $\hat{B}_i$ could be (in general) linear combinations of (tensor product) Pauli operators (see equation 15) because the basis change from {|$e_i$>} to {|$\mu_j$>}. The evolution operator eliminates the possible initial factorization between the state of the register and environment. Suppose the initial state is characterized by the tensor product of the density operator for the system and environment: $\hat{\rho}(0)_s \otimes \hat{\rho}(0)_e$. The whole evolution can be written as:

$$\hat{\rho}_s(0) \otimes \hat{\rho}_e(0) \xrightarrow{\text{Evolution: } \hat{U}} \hat{\rho}(t) = \hat{U}\left[\hat{\rho}_s(0) \otimes \hat{\rho}_e(0)\right]\hat{U}^+ \rightarrow \hat{\rho}_{rs}(t) = tr_e\{\hat{\rho}(t)\} \qquad (52)$$

$\hat{\rho}(t)$ being the density operator of the {system + environment} at time t, $\hat{\rho}_{rs}(t)$ the reduced density operator (or matrix) of the system obtained taking the partial trace with the environment states. Carrying out the calculation:

$$\hat{\rho}_{rs}(t) = tr_e\{\hat{\rho}(t)\} = \sum_i \langle\mu_i|\hat{U}\left[\hat{\rho}_s(0) \otimes |e\rangle\langle e|\right]\hat{U}^+|\mu_i\rangle =$$
$$= \sum_i \langle\mu_i|\hat{U}|e\rangle \hat{\rho}_s(0) \langle e|\hat{U}^+|\mu_i\rangle = \sum_i \hat{B}_i \hat{\rho}_s(0) \hat{B}_i^+ \qquad (52)$$

where $\hat{B}_i = \langle\mu_i|\hat{U}|e\rangle$ are operators acting on the Hilbert space of the system. Using the definition is not difficult to show the normalization condition $\sum_i \hat{B}_i^+ \hat{B}_i = I_s$ (identity for the system). The map:

$$\hat{\rho}_s(0) \mapsto E(\hat{\rho}_s(0)) = \hat{\rho}_{rs} = \sum_i \hat{B}_i \hat{\rho}_s(0) \hat{B}_i^+ \qquad (53)$$

defines a quantum operation representing the density operator evolution of the system *alone*. All the environment effect is hidden in $\hat{B}_i$, called *interaction operators*. The breaking down of $\hat{\rho}_{rs}$ in terms of $\hat{B}_i$ is called *operator-sum representation or Kraus representation*. Notice this representation in terms of $\hat{B}_i$ is not unique because it is environment basis-dependent.

The depolarizing error model applied to a qubit q and shown in equation (14), can be described now by means of the following interaction operators:

$$\hat{B}_1 = \sqrt{1-\varepsilon}\,\hat{A}_1 \qquad \hat{B}_i = \sqrt{\frac{\varepsilon}{3}}\,\hat{A}_i \quad \text{with}\quad i=2,3,4 \tag{54}$$

describing the evolution of a qubit density operator:

$$\hat{\rho}_q(0) \to \mathsf{E}(\hat{\rho}_q(0)) = \hat{\rho}_{qr}(t) = (1-\varepsilon)\,\hat{\rho}_q(0) + \frac{\varepsilon}{3}\sum_{i=2}^{4}\hat{A}_i\,\hat{\rho}_q(0)\,\hat{A}_i^{+} \tag{55}$$

Redefining the parameter $\varepsilon = 3p/4$, the reduced density matrix is:

$$\mathsf{E}\left(\hat{\rho}_q(0)\right) = \hat{\rho}_{qr}(t) = p\frac{\hat{I}}{2} + (1-p)\hat{\rho}_q(0) \tag{56}$$

Its evolution is now very transparent, showing two contributions: the untouched qubit with probability (1-p) and a completely mixed state $\hat{I}/2$ with probability p.

The correction process can be seen as the search for the inverse quantum operation $\mathsf{E}^{-1}$. In spite of the $\mathsf{E}^{-1}$ in the whole Hilbert space ($\mathcal{H}^{\otimes n}$) only exists in case of a unitary operator, it is possible to invert it taking the restriction to some special subspaces $Q \subset \mathcal{H}^{\otimes n}$. The quantum recovery operation $\mathsf{R}$ fulfils the condition:

$$\forall \rho \text{ defined in } Q, \qquad \mathsf{R}(\mathsf{E}(\rho)) \propto \rho \tag{57}$$

The recovery $\mathsf{R}$ reverses the errors represented by $\mathsf{E}$, mapping them into an operator proportional to the identity (equation 57).

The notion of detectable errors has been explicitly introduced by Knill and Laflamme [3], and can be established as: an error $\hat{B}$ is detectable by the quantum code Q if and only if

$$\forall |u\rangle, |v\rangle \in Q \quad \text{fulfilling} \quad \langle u|v\rangle = 0 \;\Rightarrow\; \langle u|\hat{B}|v\rangle = 0 \tag{58}$$

The error $\hat{B}$ transforms the Q-codewords keeping their orthogonality and being able to differentiate them. Alternatively, an error $\hat{B}$ is detectable by Q if and only if the condition $\hat{P}_Q \hat{B} \hat{P}_Q = \alpha_B \hat{P}_Q$ is satisfied for a complex constant $\alpha_B$ depending on the error $\hat{B}$. $\hat{P}_Q$ being the projector operator in Q. Without going into a full demonstration (see details in Nielsen and Chuang [3] and [27]), imagine the condition is fulfilled $\forall |\phi_{1,2}\rangle \in \mathcal{H}^{\otimes n}$, and $\hat{P}_Q|\phi_{1,2}\rangle = |u_{1,2}\rangle \in Q$. If $\langle u_1|u_2\rangle = 0$, the condition means $\langle \phi_1|\hat{P}_Q \hat{B} \hat{P}_Q|\phi_2\rangle = \langle u_1|\hat{B}|u_2\rangle =$ $\alpha_B \langle u_1 | u_2 \rangle = 0$. Consequently, a set of errors $C_Q = \{\hat{B}_i\}$ is called correctable by the code Q if and only if the set $C_Q C_Q^+ = \{\hat{B}_i \hat{B}_k^+\}$ is detectable. Equations 31 are an example of that fact.

From the general condition $\hat{P}_Q \hat{B}_i \hat{B}_k^+ \hat{P}_Q = \alpha_{ik} \hat{P}_Q$ it is possible to obtain the recovery quantum operation R. As a quantum operation it is characterized by means of the set $\{\hat{R}_i\}$ defining the map:

$$\hat{\rho} \mapsto R(\hat{\rho}) = \sum_i \hat{R}_i \hat{\rho} \hat{R}_i^+ \tag{59}$$

The matrix $\alpha_{ik}$ only depends on the error operator $\hat{B}_i \hat{B}_k^+$, and its elements are $\alpha_{ik} = \langle u | \hat{B}_i \hat{B}_k^+ | u \rangle$, $|u\rangle \in Q$. Because it is hermitian can be diagonalized, and the new set of errors $\{\hat{N}_i\}$ obtained as the appropriate linear combinations of $\{\hat{B}_i\}$, have $\{d_a\}$ as eigenvalues. For each $d_a \neq 0$ (if $d_a = 0$, $\hat{N}_a | u \rangle = 0$, $\forall | u \rangle \in Q$), a recovery quantum operation can be defined as:

$$\hat{R}_a = \frac{1}{\sqrt{d_a}} \left( \sum_{|u\rangle \in Q} |u\rangle\langle u| \right) \hat{N}_a^+ \tag{60}$$

fulfilling the condition $\hat{R}_a(\hat{N}_b) = \sqrt{d_a} \delta_{ab} \hat{I}_Q$ for the states in the code Q. Then $\hat{R}_a$ corrects $\hat{N}_a$ in Q and R corrects any linear combination of $\hat{N}_a$ errors. Notice that, strictly speaking, to have a quantum recovery operation R, it has to be extended to the whole Hilbert space (for mathematical details see Preskill [3]) that can be split as $\mathcal{H}^{\otimes n} = \left( \bigoplus_a \hat{N}_a Q \right) \oplus Q^\perp$, where $Q^\perp$ is the orthogonal complement of the code Q which is not reached acting on the code with the operators $\hat{N}_a$.

In order to implement the recovery quantum operation, the operators $\hat{R}_a$ can be written as:

$$\hat{R}_a = \frac{1}{\sqrt{d_a}} \left( \sum_{|u\rangle \in Q} |u\rangle\langle u| \right) \hat{N}_a^+ = \frac{\hat{N}_a^+}{\sqrt{d_a}} \left( \sum_{|u\rangle \in Q} \hat{N}_a |u\rangle\langle u| \hat{N}_a^+ \right) = \hat{N}_a^+ \hat{P}_{aQ} \tag{61}$$

The operator $\hat{P}_{aQ}$ projects onto the subspace $\hat{N}_a Q \subset \mathcal{H}^{\otimes n}$. Its implementation involves the projection of the corrupted state onto $\hat{N}_a Q$ according to $\hat{P}_{aQ}$, identifying the subspace $\hat{N}_a Q$ characterized by the index a, and then apply the inverse operator $\hat{N}_a^+$. To carry out the recovery process an ancilla system is introduced, characterized by the Hilbert space $\mathcal{H}_A$ and with a set of standard orthogonal states $\{|a_r\rangle\}$. Now we work in the subspace $\left( \bigoplus_a \hat{N}_a Q \right) \otimes \mathcal{H}_A$. The first step is to apply the unitary operator $\hat{V}$ (the $|a_0\rangle$ state is the initial ancilla state):

$$\hat{V} = \sum_a \hat{P}_{aQ} \otimes |a_{S_a}\rangle\langle a_0| \qquad (62)$$

The operator $\hat{V}$ is a generalization of the standard controlled-operation and will project onto $\hat{N}_a Q \otimes |a_{S_a}\rangle$. The ancilla state carries the syndrome information $S_a$ of the $\hat{N}_a$ error. Measuring the ancilla in the standard basis we obtain the state $|a_{S_a}\rangle$ and, finally, applying the operator $\hat{N}_a^+$, the error is reversed. This general process can be recognized in what it was done in figures 7 (section 10.1.7) and 9 (section 10.4.1) for the three qubit repetition code and Steane code, respectively.

**10.6 Fault tolerance in QECC**

The final mission of the decoherence control in quantum computers is a static stabilisation of the information when it is transmitted or stays in the memory, as well as a dynamic stabilization of it. We need to process the information *dynamically*, applying gates without an excessive accumulation of errors, during the time sufficient to complete the execution of the algorithm. The error correcting codes are the first step to reaching it, the second is the use of fault-tolerance techniques [28].

To implement a quantum gate, we could decode the quantum state, carry out the gate and encode the state again. This process is not advantageous since during the period of time in which the gate is put into operation, the information is unprotected. The fundamental idea of fault-tolerance is to use an encoded logic: *applying the encoded quantum gates to encoded qubits* [29], *without a previous decoding*. Nevertheless, the encoded logic, by itself, is not sufficient to assure its tolerance to failures and we will have to consider two additional aspects. In the first place, the application of encoded gates to encoded qubits can disperse the errors to others qubits within the same register as well as to other registers, until they become uncorrectable. Secondly the error correction processes are also quantum computations, which is why they can introduce new errors. We will have to make an appropriate design of encoded gates and error correction circuits to control error dispersion and accumulation. Reaching these objectives, we will make periodic encoded corrections to the qubits.

10.6.1 Error propagation

One of the frequent types of gates in the computation and error correction are the control-M (M = NOT, Z). Let us see how the CNOT gate propagates the errors. A bit-flip error in the control qubit of a gate CNOT, propagates *forward* towards the target qubit. In addition to this spread (of classic type), a phase-flip error propagates backwards, from the target qubit to the control qubit. Let us suppose a CNOT gate whose control qubit is |q> = a|0> + b|1> and a phase-flip error occurs in the target qubit (|0>+|1> in figure 15). The phase-flip error ($\hat{Z}$) propagates from the target to |q>. The bit-flip and phase-flip error propagation is shown in figure 16. Similar situations arise in gates involving several qubits, such as the Toffoli gate. If we use a code allowing a single error to be corrected in each quantum register, we define a fault-tolerant procedure as one with the property that *if an error happens in one of its components, it causes (at most) one error in each register*. The uncorrectable errors (for example in two qubits) take place with a probability $O(\varepsilon^2)$, $\varepsilon$ being the probability per qubit that some time step or gate introduces an error. This definition can be generalized into codes that correct t errors, just by demanding that no more than t errors are introduced in each register after the procedure execution.

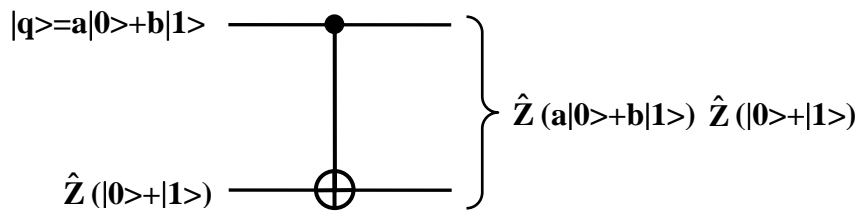

Fig. 15. Phase-flip ($\hat{Z}$) error propagation from the image qubit to the control qubit in the case of a CNOT gate connecting both qubits.

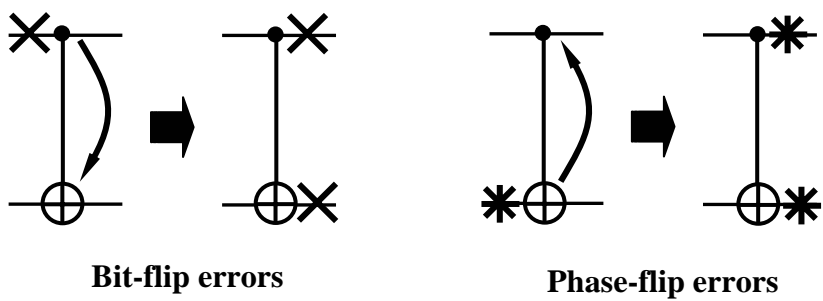

**Bit-flip errors**     **Phase-flip errors**

Fig. 16. Forward bit-flip and backwards phase-flip error propagation, due to the application of a CNOT gate.

In case that single qubits of the same register are related by a CNOT gate, the dispersion of errors could be fatal. Some gates exist, depending on the code, which can be implemented by means of a *transversal logic*, which assures its fault-tolerance. For example a CNOT gate can be transversally implemented in a [[3,1,3]] code as shown in figure 17. A bit-flip error appearing in the third qubit of the control register, propagates solely (following the arrow) to the third qubit of the target register. The CNOT gate is implemented *transversally*, by means of a procedure in which each qubit of the control register is connected to a single qubit of the target register. A transversally applied gate assures that it is fault-tolerant. An error in each register can be corrected in a later error correction step, thus avoiding its accumulation. The error probability that two uncorrectable errors occur in the control register (which would induce two errors in target qubit) behaves like $O(\varepsilon^2)$.

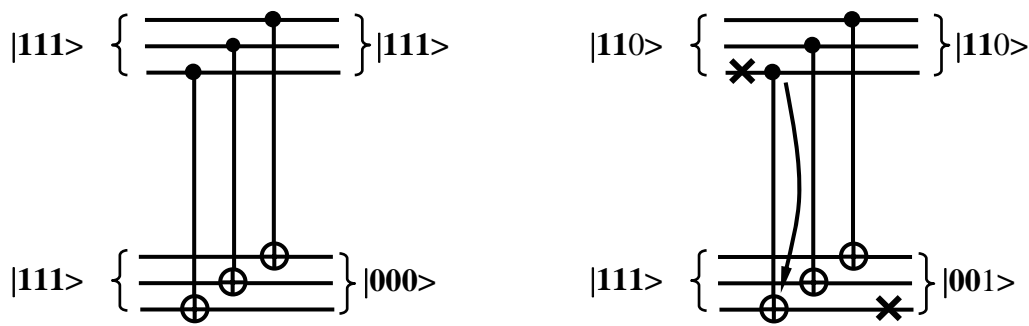

Fig. 17. Encoded CNOT gate in the [[3,1,3]] code. The left-hand piece shows a perfect transversal CNOT gate performance when the control and target registers are |111>. On the right-had side, a bit-flip error corrupts the third qubit of the control register, and is dispersed to the third qubit of the target register.

Some gates cannot be implemented in a transversal form, as would be the situation for the Hadamard rotations in [[3,1,3]], and we should design more complex circuits that could involve qubit measurements.

10.6.2 Fault-tolerant error correcting circuits

Quantum error correcting codes imply encoding, syndrome measurement and qubit correction processes using ancilla qubits, and adding new time steps and gates. In this situation the errors become more probable, with new routes appearing in the error spreading. For the QECC to be useful it is necessary for their implementation to be sufficiently robust, avoiding more errors being introduced into them than they try to eliminate, as well as being sufficiently fast.

Let us suppose that we use a QECC and the probability that a register has an error is $O(\varepsilon)$, coming up from evolution or gate errors. The definition of a fault-tolerant error correction circuit reflects the intuitive idea that it must correct more errors than are introduced by it. A quantum circuit (of a code with distance 3) is considered fault-tolerant if the probability of a register having an uncorrectable error after its execution, behaves like $O(\varepsilon^2)$. In general a quantum circuit correcting t errors is fault-tolerant if the probability of uncorrectable errors is $O(\varepsilon^{t+1})$. The tolerance to failures tries to avoid all ways in which such possibilities can take place.

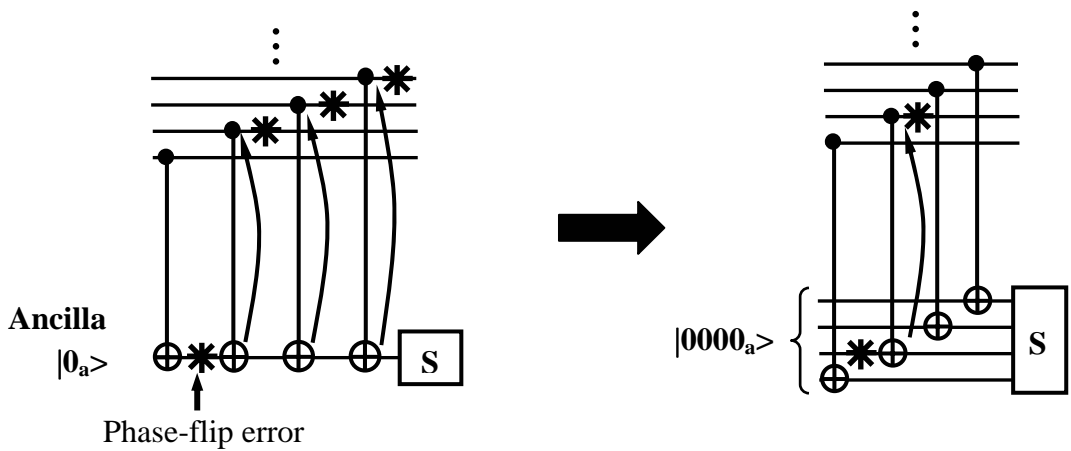

Fig. 18. Dispersion of phase-flip errors in the measurement of one bit of syndrome S. On the left, a phase-flip error propagates from an ancilla qubit until the three qubits of the upper register. This error cannot be corrected. On the right there is an equivalent transversal version. A phase-flip error in the ancilla has been propagated exclusively to one qubit of the upper register. In this case the error could be correctable in later time steps.

One of the most important steps in the correction processes are destructive and non-destructive measurements. They are used in the encoding, syndrome determination and ancilla synthesis (employed not only in error corrections, but in the implementation of encoded fault-tolerant gates), this is why its fault-tolerant accomplishment is of fundamental importance. Concerning codes with distance three, to get a fault-tolerant encoded measurement process, two conditions must be fulfilled:

a) An error in any time step of the measurement process must produce one error in each register, and
b) If an error occurs during the measurement process, the probability that the result of the measurement is incorrect must be $O(\varepsilon^2)$.

The motivation of the first condition is that an error in a register is tolerable by the code and can be corrected in a later time step. Whereas the second, reflects the fact that the measurement results could be used to correct one or several qubits within a register. If the error probability in a measurement behaves like $O(\varepsilon)$, the subsequent error correction using this result, could introduce several errors into the same register with $O(\varepsilon)$ probability, with them being uncorrectables.

The destructive measurement produces the collapse of the measured state. That is the case in the final ancilla measurement for obtaining the error syndrome. If the error probability is $O(\varepsilon)$ in each qubit, and these are not correlated, the probability of an uncorrectable error taking place (two or more errors in the register) is $O(\varepsilon^2)$. So the destructive measurements are fault-tolerant.

As well as the destructive measurements, we can measure hermitian operators non-destructively, such as the stabilizer generators, as we have already seen in figure 12. The syndrome copied into the ancilla, is known by measuring this last one destructively. Nevertheless a single ancilla state is not suitable to extract the syndrome because the circuit could spread the errors in an uncorrectable way. As an example, let us look at a piece (figure 18) of the correction circuit shown in figure 9 (measurement of the syndrome $S_1$). A phase-flip error in the ancilla qubit (we suppose in the state $|1_a\rangle$) used as the target of several CNOT gates, is propagated to three qubits of the upper register. If the phase-flip error probability is $O(\varepsilon)$, the propagation has introduced three errors affecting the upper register with the same probability, being uncorrectable in later corrections (left-hand side in figure 18). This behaviour appears because the same ancilla qubit is the target of all the CNOT gates. We can solve the problem, replacing the ancilla qubit by four, so that if there is an error in one these single qubits, they propagate to a single qubit in the control register (the right-hand side of figure 18). Nevertheless this operation does not solve completely the problem. Let us see why in the following example.

Suppose we use the Steane [[7,1,3]] code to get $|0_E\rangle$ and $|1_E\rangle$, and a bit-flip error occurs in the seventh qubit, represented by $\hat{X}_7 \equiv \hat{X}_e$ (e = (0000001)). To obtain one bit of syndrome (for example, measuring the $\hat{Z}_{0001111}$ generator shown in figure 18), we use an ancilla state $|0000_a\rangle$ and four CNOT gates are applied involving four qubits of the encoded control register and four ancilla target qubits. The process would be as follows:

$$\hat{X}_7 \left\{ a|0_E\rangle + b|1_E\rangle \right\} |0000_a\rangle =$$

$$\left[ a\{|0000001\rangle + ...\} + b\{|1111110\rangle + ...\} \right] |0000_a\rangle \xrightarrow{\text{four CNOT gates}}$$

$$\rightarrow a\left(\hat{X}_7 |0_E\rangle\right) |0001_a\rangle + b\left(\hat{X}_7 |1_E\rangle\right) |1110_a\rangle$$

(63)

Measuring the ancilla destructively, we will find two states $|0001_a\rangle$ or $|1110_a\rangle$ bringing about a collapse of the whole state to $\hat{X}_7|0_E\rangle$ or to $\hat{X}_7|1_E\rangle$, with probabilities $|a|^2$ and $|b|^2$ respectively. The initial qubit coherence has been destroyed, acquiring certain information about it. To cope with the problem we need to synthesize special ancilla states and design appropriate recovery circuits.

10.6.3 Ancilla states

Ancilla states are involved in syndrome measurement as well as intermediate states in fault-tolerant encoding and gates; therefore the design of the ancillas is an important aspect in the fault-tolerant computation. In QECC we require an ancilla state to copy the error syndrome on to it and that we can measure destructively without losing the qubit coherence and without introducing too many new errors. The syndrome measurement does not have to reveal anything about the state of the information qubit.

DiVincenzo and Shor propose using a special ancilla state $|a_{Shor}\rangle$ synthesized from the Schrödinger cat-state ($|0000\rangle + |1111\rangle$), and Hadamard rotating all qubits to obtain an entangled state of equal weighted even parity registers:

$$|a_{Shor}\rangle = \frac{1}{\sqrt{8}}\{|0000\rangle + |0011\rangle + |0101\rangle + |1001\rangle + |1010\rangle + |1100\rangle + |0110\rangle + |1111\rangle\} \quad (64)$$

In order to find each bit of syndrome, one Shor ancilla state is needed, by copying the syndrome on to it by the appropriate CNOT gates and being, at the end, destructively measured. But why is this way of copying the information advantageous? After applying the CNOT gates and its measurement, the ancilla state (carrying the information of the error) is not entangled with the qubit, so the ancilla measurement does not collapse the qubit state. The ancilla collapse takes place randomly in one of its even parity registers, preventing information from being obtained about the qubit. If the parity of the ancilla register has changed, the syndrome bit is 1, in another case it will be 0. The parity only reveals the bit of syndrome and nothing about the qubit coefficients. For example, using Steane's code, we need three bits of syndrome to store the bit-flip error information and three more for phase-flips, altogether six Shor ancilla states.

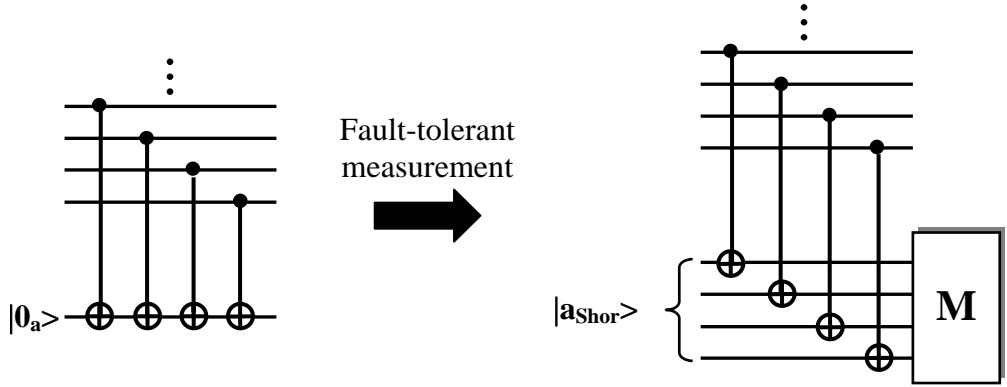

Fig. 19. Fault-tolerant circuit for the measurement of one bit of syndrome corresponding to the $\hat{I}\hat{I}\hat{I}\hat{Z}\hat{Z}\hat{Z}\hat{Z}$ generator of the [[7,1,3]] code. The different CNOT gates are applied transversally connecting different qubits within each register.

In the previous example an $\hat{X}_7 \equiv \hat{X}_e$ error took place with e = (0000001); after applying the four CNOT gates (involved in the $\hat{Z}_{0001111}$ generator) and the measurement, the ancilla is not entangled with information qubit, and the state is:

$$\hat{X}_e\{a|0_E\rangle + b|1_E\rangle\} \otimes \hat{P}_M |(h_{34}e_4, h_{35}e_5, h_{36}e_6, h_{37}e_7) \oplus a_{Shor}\rangle \quad (65)$$

Where $h_{ij} = (H_{[7,4,3]})_{ij}$ (parity check matrix for the [7,4,3] code, equation 42) and the operator $\hat{P}_M$ represent a projective ancilla measurement to get a single four qubit register, whose parity will provide the bit of syndrome. The circuit implementing the fault-tolerant measurement of the $\hat{Z}_{0001111}$ generator is shown in figure 19. Once the error has been identified, it is corrected applying the inverse operator. Similarly we could copy the syndromes of the generators involving $\hat{X}$ operators, just making Hadamard rotations before and after the CNOT gates and using the previous ancilla states.

Another kind of ancilla has been proposed by Steane. It involves an entangled state whose codewords correspond to those of [7,4,3] classic Hamming code:

$$|a_{Steane}\rangle = \frac{1}{\sqrt{2}}\{|0_E\rangle + |1_E\rangle\} = \frac{1}{4}\sum_{v \in C}|v\rangle \qquad (66)$$

10.6.4 Synthesis and ancilla verification

The ancillas are quantum states whose synthesis involves noisy circuits, and letting them interact with the information qubit, can propagate errors. So it is essential to take special care preparing high quality ancilla states.

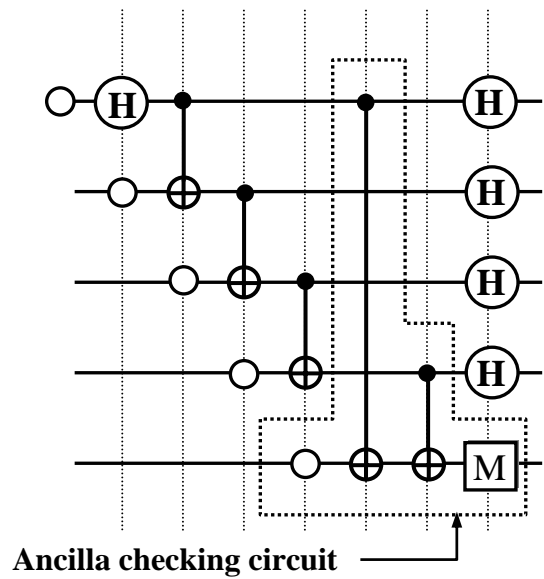

**Ancilla checking circuit**

Fig. 20. Circuit to synthesize Shor ancilla state. To check the ancilla quality a fifth qubit and two new CNOT gates are added with a final measurement M. Open circles on the left represent |0> qubit states.

Consequently, to synthesize the Shor ancilla, the simple circuit shown in figure 20, can be used. The first Hadamard rotation on the left along with the CNOT gates, creates a cat-state (|0000> + |1111>), that is transversally Hadamard rotated to obtain the final state. If an error occurs in certain locations of the circuit, it can be converted into two or more errors by the CNOT gates, dispersing them to the information qubit during the syndrome measurement. A bit-flip error in the region of the first three CNOT gates, could be propagated to two (or more) errors, that are transformed into two (or more) phase-flip errors in the final Hadamard rotations, reaching the information qubit by backwards propagation (see section 10.6.1). We arrive at an uncomfortable situation since we would have to use another error correction for the ancilla. Trying to control the ancilla bit-flip error contamination, we add a fifth qubit and two CNOT gates whose control is the first and fourth qubits and the target is the fifth. In fact any two qubits could be used instead of the first and fourth, for instance the second and fourth or third and fourth. If a single bit-flip error occur, the first and fourth qubit will have different values; acquiring the fifth qubit the value 1. Supposing that the destructive measurement of the fifth qubit is error free, a result 1 would imply discarding the ancilla and preparing a new one. If the measurement detects a 0, we proceed with the syndrome measurement from information qubit, with the security that the appearance of two phase-flip errors in the ancilla final state perform as $O(\varepsilon^2)$.

10.6.5 Syndrome verification

As we previously indicated, a bit-flip error in the ancilla synthesis circuit can be propagated as phase-flip errors on the information qubit. This possibility is controlled verifying the ancilla state by means of an ancilla checking circuit. Moreover, if the ancilla has a phase-flip error, the $\hat{H}$ gates (figure 20) transform it into a bit-flip error providing a wrong syndrome. As well as the errors in ancilla, the circuit for the syndrome extraction can also introduce errors (with probability $\varepsilon$) providing an incorrect syndrome that will contaminate the qubit, if this syndrome is used for the correction. This results in two or more unrecoverable errors with a $O(\varepsilon)$ probability.

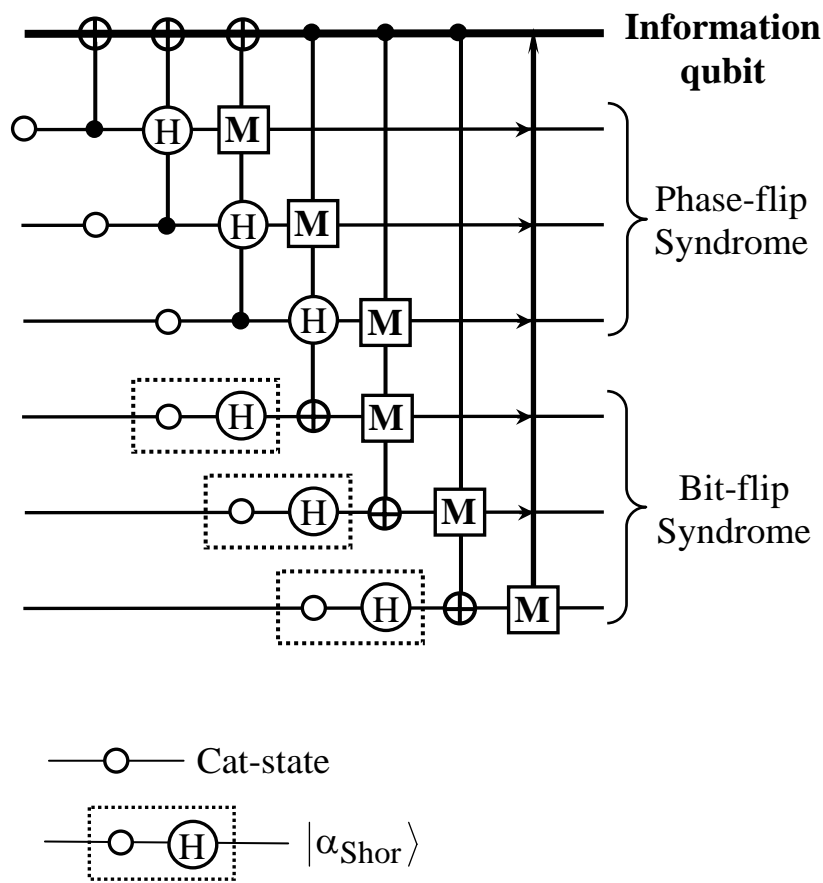

Fig. 21. Quantum circuit extracting six bits of syndrome, characterising any quantum error.

We must make sure that the syndrome is correct, for example repeating it several times. If the syndrome indicates that there is no error, we could repeat it to verify the value. If both are equal we do nothing. If the syndrome indicates an error, we repeat the syndrome and if we obtain the same one, it will be used to correct the qubit. It is possible that both syndromes are erroneous, whereas the information is correct, but this situation has a probability $O(\varepsilon^2)$. If both, first and second syndromes do not agree (due to an error in the information and another one in the syndrome; situation with a probability $O(\varepsilon^2)$), we can obtain a third one, choosing the syndrome repeated twice. In the case of three different syndromes, we can continue to calculate new syndromes until two of them agree or (more economically), we do not take any action waiting for the next recovery step. Some variations of this strategy can be raised that optimise the method.

The circuit measuring the six bits of syndrome, three for bit-flip and three for phase-flip errors, is shown in figure 21. In fact, each one of the CNOT gates corresponds to four of them, connecting ancilla qubits with the appropriate ones of the information qubit register, according to the classic [7,4,3] code parity check matrix of equation 42. The open circles on the left represent cat-states. The upper part of the circuit (figure 21) detects phase-flip errors and is made up using the equivalence shown in figure 14. The lower piece of the circuit detects the bit-flip errors and the states inside the dotted boxes represent $|a_{Shor}\rangle$ states. Gates M are destructive ancilla measurements.

10.6.6 Numerical simulation of an error correction

Using the depolarising error model, we have simulated the qubit error correction encoded by means of the Steane quantum code [[7,1,3]] [30]. In order to show the advantages of the fault-tolerant methods, two schemes for the syndrome extraction have been used. Firstly, by means of a non fault-tolerant ancilla and secondly using Shor's.

The specific qubit $|q\rangle = (|0\rangle + |1\rangle)/2^{1/2}$ is encoded as $|q_E\rangle$, via the circuit shown in figure 22, and subsequently an error correction is applied. Although the encoding circuit is not fault-tolerant, it is used as a reference circuit creating the initial noisy state. As an example of the simulation a fixed gate error probability of $\gamma = 0.001$ has taken, calculating the final state fidelity as a function of $\varepsilon$ (free evolution error probability). In all the cases simulated, the encoding includes evolution as well as gate errors, with an error probability $O(\varepsilon,\gamma)$. These could propagate errors in two or more qubits with the same probability because the encoding circuit represented in figure 22 is not fault-tolerant.

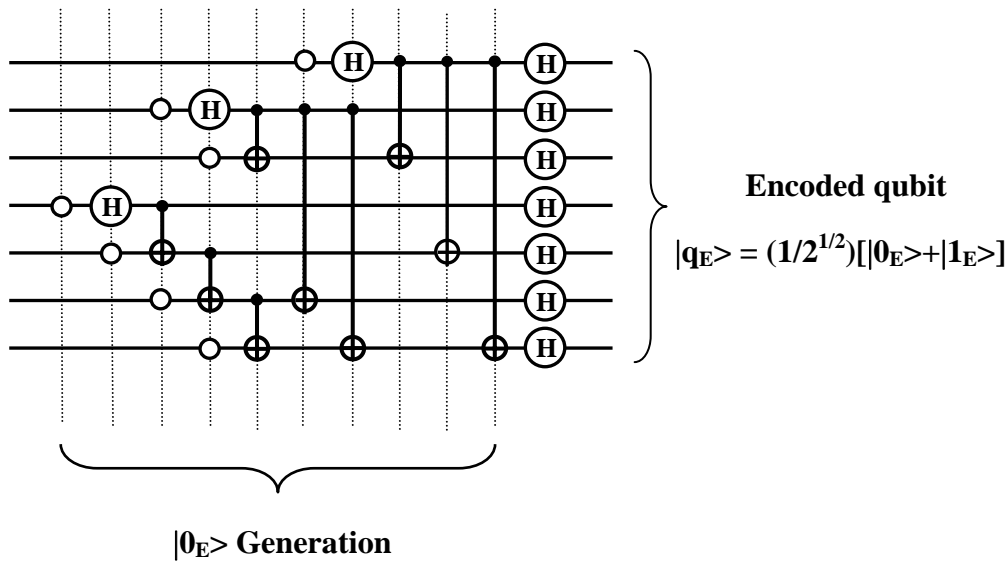

Encoded qubit

$|q_E\rangle = (1/2^{1/2})[|0_E\rangle+|1_E\rangle]$

$|0_E\rangle$ Generation

Fig. 22. Circuit encoding the qubit $|q\rangle = (|0\rangle +|1\rangle)/2^{1/2}$ by means of a [[7,1,3]]. Open circles on the left represent $|0\rangle$ states. The first part of the circuit generates a $|0_E\rangle$ state, which is transformed into the final qubit state by means of a transversal Hadamard rotation.

In the first simulation case, a simple three qubit $|000_a\rangle$ ancilla state is used, on to which the error syndrome is copied. If the correction circuit worked perfectly, it would correct all the errors of weight one, which is why we would hope that the fidelity would behave as $F_E(\varepsilon,\gamma) = 1 - O(\varepsilon^2, \gamma^2)$. Nevertheless, since the encoding circuit is not fault-tolerant, a linear term appears in $F_E(\varepsilon,\gamma)$. We fit the simulation results (figure 23) for $F_E(\varepsilon,\gamma=0.001)$ to a polynomial of degree 3 in $\varepsilon$, providing a linear term of $-2.26\,\varepsilon$ (undergoing only small variations when the degree of the polynomial increase). Actually, the correction process is a quantum computation and, therefore noisy. If we also introduce errors into the correcting process step, the result obtained for $F_E(\varepsilon,\gamma=0.001)$ (figure 23) has a linear term $-77.47\,\varepsilon$, and the fidelity quickly decreases as $\varepsilon$ increases.

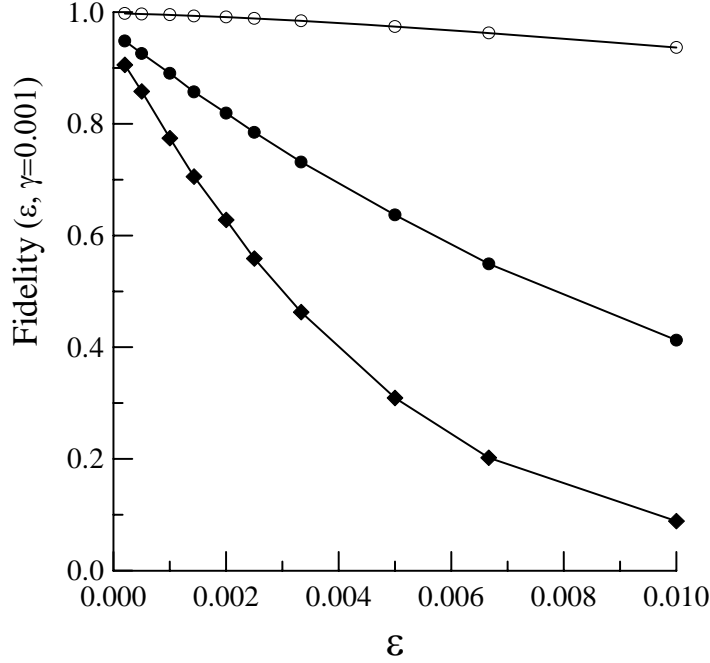

Fig. 23. Fidelity versus $\varepsilon$ for $\gamma=0.001$. The encoding circuit always is affected by gate and free evolution errors. The error correction step could be noisy (with errors) or perfect (without errors). The qubit $|q_E\rangle$ encoded by means of the circuit shown in figure 22, is later corrected with the following methods: o simple ancilla $|000_a\rangle$ with perfect correction step; • simple ancilla $|000_a\rangle$ with a noisy correction step and ♦ fault-tolerant method using Shor's ancilla,

Instead of using a simple ancilla whose initial state is $|000_a\rangle$, we can use $|a_{Shor}\rangle$, repeating the syndrome three times before correcting the qubit. In this way we hope to improve the previous results, since the complete correcting method is now fault-tolerant. The simulation produces a fidelity (see figure 23) that, surprisingly seems worse than that obtained with the simple ancilla, displaying a linear term $-184.2\,\varepsilon$.

So, where is the advantage in using a fault-tolerant error correction? We must try to find the answer in the error accumulation over the time. Whereas the appearance of one or two errors in $|q_E\rangle$ provides zero fidelity, both situations are not equally pernicious. In the second case the encoded qubit state is not recoverable whereas in the first it is. To appreciate the advantage of using a fault-tolerant ancilla we can make a simulation for the error correction of the qubit perfectly encoded (without error) and sent through the channel with only free evolution noise of probability $\varepsilon$. The noisy correction process always includes evolution as well as gate error. When the gate and evolution errors are sufficiently small ($\varepsilon = 10^{-4}$ and $\gamma = 2\,10^{-4}$ in figure 24), the Shor ancilla state with three syndromes avoids the pernicious error accumulation over the time. For the results shown in figure 24, beyond 140 time steps, the fidelity obtained with a fault-tolerant method is better than that obtained with the simple ancilla.

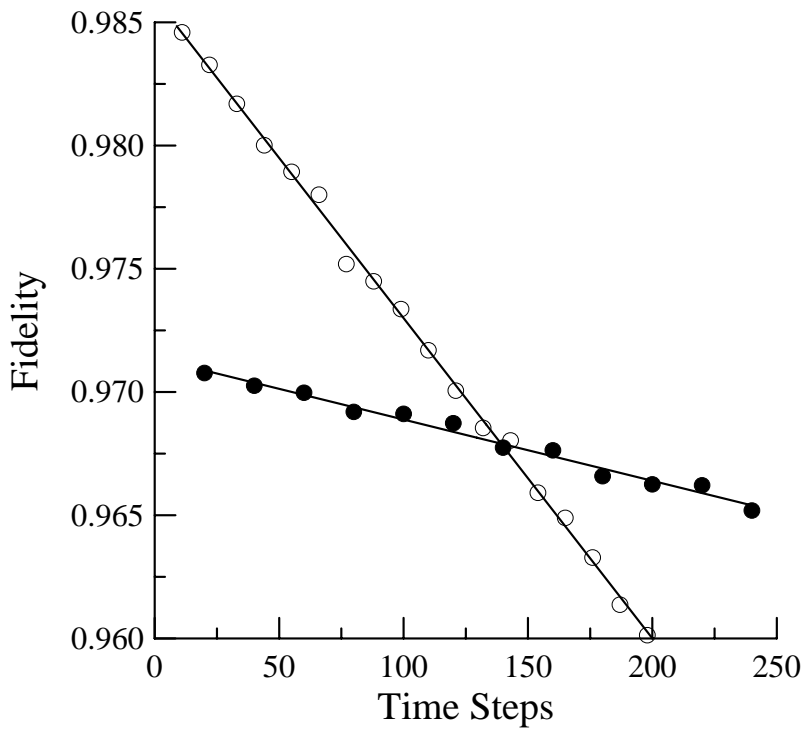

Fig. 24. Fidelity versus time steps for $\varepsilon = 10^{-4}$ and $\gamma = 2 \cdot 10^{-4}$. The curves correspond to a perfect qubit encoding and noisy correction through a free evolution noisy channel of probability $\varepsilon$ by means of two methods: o simple ancilla $|000_a\rangle$ and • fault-tolerant Shor

More elaborate fault-tolerant strategies, the use of parallelised ancilla states or simpler circuits of interaction ancilla-qubit could be more advantageous with respect the use of simple ancillas.

**10.7 Concatenated quantum codes and threshold theorem**

We already have in our hands the fundamental bricks to carry out a quantum computation robust to failures. Given a quantum circuit, we encode each qubit by means of a suitable code correcting $t = \lfloor (d-1)/2 \rfloor$ errors, using an encoded fault-tolerant logic that control the dispersion and error accumulation. After each encoded gate, we make a correction in each register using a fault-tolerant error correction circuit. This scheme seems to permit a computation for a time long enough to implement any quantum algorithm. We only need to choose a code with t large enough, since the probability that an uncorrectable error appears in (t+1) qubits is $O(\eta^{t+1})$ ($\eta$ being the error probability per non-encoded qubit and time step). Nevertheless an additional problem appears. In order to implement codes with increasing values of t, more complex circuits and greater number of qubits are needed, so the small probability of error $O(\eta^{t+1})$, begins to be important. The recovery circuit can introduce more errors than those it eliminates.

Shor [31] studied this situation in the case of Reed-Muller codes, concluding that to make T time steps with a small probability of error, it was necessary for the gate or time step error to behave like $O(1/\log^4 T)$. The dependence of the tolerable error with the number of time steps seems to prevent long computations. We need codes whose t values increase more quickly than the complexity of their recovery circuits, these codes are the *concatenated quantum codes* [32].

The concatenated codes use an encoding hierarchy. A possible construction scheme is as follows. Each qubit is encoded with a quantum code $Q_1 = [[n_1,k,d_1]]$ (first encoding level). The resulting qubits encoded in the first encoding level are encoded again with the quantum code $Q_2 = [[n_2,1,d_2]]$ (second encoding level), and so on. We could say that a concatenated code is a code within another one. The resulting code has the parameters $[[n_1n_2,k,d \geq d_1d_2]]$. A particular case of concatenated code is the Shor code. It was created by a repetition code correcting one bit-flip error $[[3,1,3]]$ whose base is $\{|000>, |111>\}$, concatenated with a later encoding of each one of the encoded qubits by means of a code that corrects one phase-flip error $[[3,1,3]]$ whose base is $\{(|0> + |1>)^{\otimes 3}, (|0> - |1>)^{\otimes 3}\}$. Although both codes have distance 3 for the errors that they correct, with respect to the set of both types of error, they have distance 1. Therefore, the resulting code is $[[3^2,1,d = 3 > 1]]$, with a distance strictly greater than the product of their distances, which is why it is capable of correcting any single quantum error.

In the case of using the same code $[[n,1,d]]$ in all the hierarchy, after L levels of concatenation (or encoding) we obtain the code $[[n^L,1, d \geq d^L]]$. So that the code can recover correctly, there must be fewer than (t+1) errors (if d = 2t+1) in the first level. The error probability in the first level $P^{(1)}$, is bounded by:

$$P^{(1)} = \sum_{i=t+1}^{n} \binom{n}{i} (1-\eta)^{n-i} \eta^i \leq \binom{n}{t+1} \eta^{t+1} \qquad (67)$$

$\eta$ being the error probability of each qubit. If t=1, $P^{(1)} \leq \binom{n}{2} \eta^2 = C \eta^2$. Likewise, the failure probability in the second level fulfils $P^{(2)} \leq C (C\eta^2)^2$ and when the L concatenation level is reached, $P^{(L)} \leq (1/C)(C\eta)^{2^L}$. If $\eta < 1/C = \eta_{th}$ (error threshold), the error probability of the concatenated code can become as small as we want adding so many levels of concatenation as is necessary. For the Steane code $[[7,1,3]]$, $\eta_{th} = 1/21$. Although the value found for $\eta_{th}$ shows the method for obtaining the threshold, its value is not real. More elaborate treatments provide $\eta_{th} \sim 6 \cdot 10^{-4}$ for the gate ($\eta_g$) and free evolution ($\eta_e$) error thresholds [28]. For errors ($\eta_g, \eta_e$) < $\eta_{th}$, given a circuit, another polynomial equivalent in size to the previous one can be found that can make a sufficiently long computation. This it is in essence the *threshold theorem* [33] for the quantum computation.

By means of the previous error model (depolarizing error channel, section 6 and 10.6.6), it is possible to make a first estimation of the computation threshold when L=1 (no concatenation is used). Considering that $\eta = \varepsilon \sim \gamma$, we compare the uncorrectable error probability in different cases when the qubit $(|0> + |1>)/2^{1/2}$ is sent through the noisy channel: (1) non-encoded qubit, (2) perfectly encoded qubit and corrected by means of a simple ancilla and (3) perfectly encoded qubit and corrected using a fault-tolerant Shors's method. In the first case, the uncorrectable error probability after t time steps is $P_1(\eta, t) = 1-(1-2\eta/3)^t$, because $\hat{Y}$ and $\hat{Z}$ errors (but not $\hat{X}$) produce zero fidelity. The simple ancilla and Shor's method takes 12 and 20 time steps [30], respectively, to carry out the error correction after one time step of free evolution. Therefore, the probabilities $P_1(\eta, 12)$ and $P_1(\eta, 20)$ are compared with the uncorrectable error probability obtained with methods (2) and (3). The results appearing in the figure 25 show a quasi-linear behaviour for $P_2(\eta, 12)$ (simple ancilla) and a complete quadratic behaviour $P_3(\eta, 20) = a \eta^2$ (with a = 19151.6) when Shor's fault-tolerant method is employed. There exist a clear crossing between $P_3(\eta, 20)$ and the line $P_1(\eta, 20) \sim 40\eta/3$ at $\eta = 40/3a = 7 \cdot 10^{-4}$. So when $\eta < 7 \cdot 10^{-4}$, a clear benefit is obtained with the method (3) compared to the non-encoded. This value is very close to Preskill's threshold [28]. A stronger fault-tolerant threshold can be infer as $\eta_{th} = 1/a = 5.2 \cdot 10^{-5}$. If $\eta < 5.2 \cdot 10^{-5}$ the error accumulation originate in the free qubit evolution is avoided to a great extent. Increasing L, the threshold would, certainly, decrease. Recently, Reichardt [34] have used the same [[7,1,3]] quantum code and the depolarizing error model to estimate the threshold but *without memory errors*, providing a smaller threshold ($9 \cdot 10^{-3}$) than the present one.

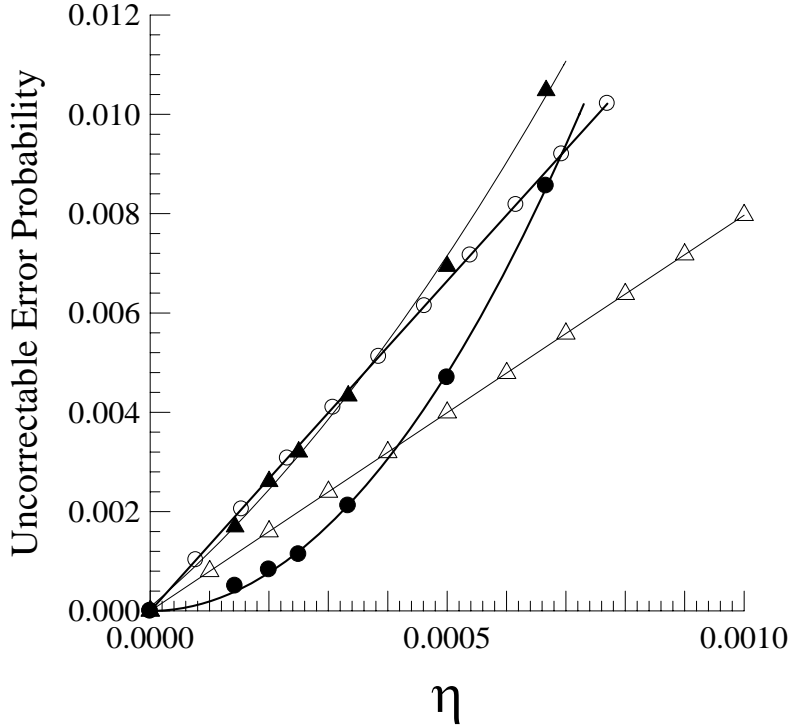

Fig. 25. Non-correctable error probability comparison between: △ $P_1(\eta,12)$, ○ $P_1(\eta,20)$, ▲ $P_2(\eta,12)$ simple ancilla and ● $P_3(\eta,20)$ fault-tolerant Shor's method.

Although the threshold theorem depends strongly on external considerations such as the error model, it demonstrates that under certain circumstances an imperfect logic does not impose a fundamental limitation for the operation of the quantum computers.

## 11. Summary

We have reviewed the fundamental ideas to control the decoherence in a quantum computer, particularly the error correcting codes. The appearance of concatenated quantum error correcting codes has provided the first victory in the decoherence control even when imperfect devices are used. Furthermore, with a simple depolarizing error model, we have been able to estimate the memory threshold ($5.2 \cdot 10^{-5}$) below which it is possible to greatly stabilize a qubit in the quantum memory. Its value is not as important as it is its own existence. In addition, it is possible to conjecture that the threshold to process the quantum information dynamically, i.e. applying quantum gates, would decrease this threshold in a factor less than ten. These values are technologically achievable, so the initial down-heartedness about the possibility of making sufficiently long computations has been overcome.

At the moment the correction circuits seem to be somewhat complex and expensive to be experimentally implemented, and it will be necessary to develop more simplified methods without losing its effectiveness. In this sense, techniques such as the decoherence free subspaces, seems to be a good way to reach these objectives.

**Acknowledgements**

The author thanks the referee's comments to the original version of this paper.